\documentclass[aps,prc,floatfix,showpacs,twocolumn,superscriptaddress,10pt]{revtex4-1}

\usepackage{dcolumn}
\usepackage{bm}
\usepackage{color}
\usepackage{amssymb}
\usepackage{amsmath}
\usepackage{graphicx}
\usepackage{amsfonts}
\usepackage{slashed}
\usepackage{pstricks}
\usepackage{float}
\usepackage{hyperref}
\usepackage{array}
\allowdisplaybreaks
\usepackage{dcolumn}
\usepackage{epsf}
\usepackage{float}
\usepackage{epstopdf}

\usepackage{gensymb}
\usepackage{mathtools}

\newcommand{\ket}[1]{| #1 \rangle}
\newcommand{\bra}[1]{\langle #1 |}
\newcommand{\expected}[1]{ \langle #1 \rangle}
\newcommand{\product}[2]{\langle #1 | #2 \rangle}

\newcommand{\bvec}[1]{\boldsymbol {#1}}
\newcommand{\bhat}[1]{{\boldsymbol {\hat {#1}}}}

\begin{document}
\title{Quantum Monte Carlo formalism for dynamical pions and nucleons}

\author{Lucas Madeira}
\email{lucas.madeira@asu.edu}
\affiliation{Department of Physics, Arizona State 
University, Tempe, Arizona 85287, USA}

\author{Alessandro Lovato}
\affiliation{INFN-TIFPA, Trento Institute for Fundamental Physics and Applications, 38123 Trento, Italy}
\affiliation{Physics Division, Argonne National Laboratory, Argonne, Illinois 60439, USA}

\author{Francesco Pederiva}
\affiliation{INFN-TIFPA, Trento Institute for Fundamental Physics and Applications, 38123 Trento, Italy}
\affiliation{Dipartimento di Fisica, University of Trento, via Sommarive 14, I-38123 Povo, Trento, Italy}

\author{Kevin E. Schmidt}
\affiliation{Department of Physics, Arizona State 
University, Tempe, Arizona 85287, USA}

\date{\today}

\begin{abstract}
In most simulations of nonrelativistic nuclear systems, the wave functions found solving
the many-body Schr\"odinger equations describe the quantum-mechanical amplitudes of
the nucleonic degrees of freedom. In those simulations the pionic contributions are
encoded in nuclear potentials and electroweak currents, and they determine
the low-momentum behavior. In this work we present an alternative quantum Monte Carlo formalism in
which both relativistic pions and nonrelativistic nucleons are explicitly included in the quantum-mechanical 
states of the system.
We report the renormalization of the nucleon mass
as a function of the momentum cutoff, an Euclidean time density
correlation function that deals with the short-time nucleon diffusion, and the pion cloud density and
momentum distributions. In the two-nucleon sector we show that the interaction of two static nucleons
at large distances reduces to the one-pion exchange potential, and we fit the low-energy constants of the
contact interactions to reproduce the binding energy of the deuteron and two neutrons in finite volumes. 
We show that the method can be readily applied to light-nuclei.
\end{abstract}

\maketitle

\section{Introduction}
Modern nuclear theory is characterized by a series of attempts to rigorously bridge the gap between quarks and gluons, the degrees of freedom of quantum chromodynamics (QCD),
and the confined phase in which massive particles such as mesons and baryons can
be regarded as the constituents of matter.
Nuclear effective field theories (EFTs) are employed to connect QCD to low-energy nuclear observables. EFTs exploit the separation between the ``hard'' ($M$, typically the nucleon mass) and ``soft'' ($Q$, typically the exchanged momentum) momentum scales. The active degrees of freedom at soft scales are hadrons whose interactions are consistent with QCD. Effective potentials and currents are derived in a systematic expansion in $Q/M$ from the most general Lagrangian constrained by the QCD symmetries. Chiral-EFT, which is
best suited to describe processes characterized by $Q \simeq m_\pi$, exploits the (approximate) chiral symmetry of QCD and its pattern of spontaneous symmetry breaking to derive consistent nuclear potentials and currents, and to estimate their uncertainties~\cite{Epelbaum:2008ga,Machleidt:2011zz}. 

Potentials and electroweak currents derived within chiral-EFT are
the main input to \textit{ab initio} many-body methods that are aimed at
solving the many-body Schr\"odinger equation associated with the nuclear
Hamiltonian~\cite{Barrett:2013nh,Epelbaum:2011md,Hagen:2013nca,Hergert:2015awm,Carbone:2013eqa,Carlson:2014vla}.
These schemes rely on the assumption that processes like the one meson
exchange are well approximated by an instantaneous interaction,
and that the meson degrees of freedom can be
integrated out and their contribution is encoded in nuclear potentials and
electroweak currents, determining their low-momentum behavior.  Not much
attention has been devoted so far to the development of techniques capable
of including mesonic degrees of freedom in these many-body calculations. There are several
reasons for this choice. The main one is that effects arising from not
assuming an instantaneous interaction are believed to be unessential
for the derivation of nuclear potentials. Without such assumption,
many-body interactions would automatically be generated already at
leading order when integrating out the meson fields. The fact that, when
neglecting dynamical effects in the meson fields, three- and many-body
interactions appear at next-to-next-to leading order (N2LO) suggests that
such effects can be considered to be sub-leading at any order.  However,
such assumptions have never been rigorously tested in
an \textit{ab initio} scheme
for a many-nucleon system.
This work devises a formalism in which testing these assumptions
is straightforward.

Even if few-nucleon sector calculations show that 
instantaneous pion interactions are justified, our approach 
can still be useful 
to compute quantities unavailable to other methods.
One advantage of our technique is
the ability to directly measure the pion degrees of freedom.
For example,
in theories where the pions are
integrated out
current operators
need to have the pion contributions calculated from the underlying 
theory
to give two- or more-body contributions. These pion contributions 
are
immediate in the present work.
Treating pions as dynamical degrees of freedom
allows us, for example, to tackle
the pion-production region in electron- and neutrino-nucleus scattering, 
which is unaccessible by standard nuclear many-body methods in which pion 
degrees of freedom are implicit.

Another possible usage of our approach, including the case
where the
instantaneous pion interactions approximation is valid, is
as a computational tool.
For example, in electronic structure calculations, methods such as
Car-Parrinello \cite{car85} are often used.
These solve the Born-Oppenheimer
electronic ground-state, i.e., the instantaneous electron 
approximation,
by promoting the electronic degrees of freedom to be dynamic 
variables,
and both the ions and the electronic degrees of freedom are 
included
in the time-dependent solutions. 
Similarly, even if the pion degrees
of freedom are well described by an instantaneous
Born-Oppenheimer approximation,
our scheme would be an efficient way to solve
for this Born-Oppenheimer state.

Most of the progress to account for explicit pions into nuclear EFT has been made so far by using lattice methods. While the inclusion of pion fields into the Lagrangians is straightforward, dynamical pions bring noise and sign problems in lattice Monte Carlo calculations \cite{hjo17}. One alternative approach is to use static pion auxiliary fields \cite{bor07,lee09}, where time derivatives are neglected, and thus pions couple to nucleons only through spatial derivatives. Since these pion fields are instantaneous, this eliminates the self-energy diagrams responsible for mass renormalization.
It is noteworthy to point out that there is a condensed matter analog to the axial-vector coupling between one nucleon and the pions, the polaron \cite{fey55}. However, the coupling between the electron and the phonons is scalar, and the bosonic degrees of freedom can be integrated out explicitly. Quantum Monte Carlo (QMC) methods have been successful at tackling both problems \cite{car92}.

In this paper, we devise a QMC framework in which both relativistic pions and nonrelativistic nucleons are explicitly included in the quantum-mechanical states of the system. From a given order chiral-EFT Lagrangian, the corresponding Hamiltonian is derived, and the pion fields are expressed in the Schr\"odinger representation. The nuclear structure problem is written in terms of the modes of the relativistic pion field, and of the position and spin-isospin
degrees of freedom
of the nucleons. QMC techniques are employed to accurately solve the corresponding Schr\"odinger equation, which is equivalent to summing all Feynman diagrams originating from a given order of the chiral-EFT Lagrangian. Resummation techniques are already employed in chiral-EFT. The nucleon-nucleon (NN) system at low angular momenta is characterized by a shallow bound state, the deuteron, and large scattering lengths, which prevents the applicability of standard chiral perturbation theory. Weinberg suggested to use perturbation theory to calculate the irreducible diagrams defining the NN potential, and apply it in a scattering equation to obtain the NN amplitude \cite{wei90}. Solving the scattering equation corresponds to summing all diagrams with purely nucleonic intermediate states  \cite{Machleidt:2011zz}. Diagrammatic resummation in chiral-EFT is also needed to describe resonances in pion-pion scattering that cannot be obtained in perturbation theory to any finite order \cite{nie00}.

Before moving to larger systems, there are several nontrivial questions arising when including pions in a QMC calculation, that need to be addressed already for the one and two nucleon cases.
One of the major issues is the assessment of finite size effects, since our calculations are necessarily limited to nucleons and pions lying in a box of side $L$ with periodic boundary conditions. This naturally introduces an infrared cutoff dependence which, together with
the ultraviolet cutoff we employ,
affects the pion-nucleon interaction.

In the single-nucleon sector, we study the energy-shift of the nucleon mass as a function of the momentum cutoff.
We also compute the pion cloud density and momentum distributions. 
In the NN sector, we first verified that our results for two static nucleons correctly reduce to the one-pion exchange potential at sufficiently large separation distance. We then fit the low-energy constants associated to the contact terms of the leading-order (LO) chiral-EFT Lagrangian to describe the deuteron and two neutrons in a finite volume.

This work is structured as follows. In Sec.~\ref{sec:methods}
we introduce our methodology, providing the explicit expressions
for the Lagrangian and Hamiltonian densities we use throughout the
manuscript. The Schr\"odinger representation of the pion field is also
presented, along with related quantities, such as the pion cloud density
and the total charge of the system. Section~\ref{sec:qmc0} describes
the ingredients required for the quantum Monte Carlo calculations.
These are the Hamiltonian for a fixed number $A$ of nucleons interacting with
the pions, the form for a trial wave function for these systems,
and the modifications to the propagator sampling needed.
We present an efficient procedure to optimize our trial wave functions,
and a method to calculate observables that do not commute with the
Hamiltonian. In Sec.~\ref{sec:results} we present our results, and in
Sec.~\ref{sec:summary} we summarize our work and give a brief outlook. In
the Appendices we list the conventions adopted in this manuscript,
we present a nonrelativistic calculation of the nucleon self-energy at
leading order,
and we investigate a different choice for the contact interactions.

\section{General formalism}
\label{sec:methods}

\subsection{Chiral Lagrangian}
\label{sec:lagrangian}

The heavy baryon leading-order chiral Lagrangian density in which only nucleon and pion degrees of freedom are included reads~\cite{Machleidt:2011zz}
\begin{align}
\mathcal{L}_0 &=
\frac{1}{2}\partial_\mu \pi_i \partial^\mu \pi_i
- \frac{1}{2} m_\pi^2 \pi_i \pi_i + N^\dagger \Big[ i\partial_0 +\frac{\nabla^2}{2M_0}\nonumber\\
&- \frac{1}{4f_\pi^2}\epsilon_{ijk} \tau_i \pi_j \partial_0\pi_k 
-\frac{g_A}{2f_\pi} \tau_i \sigma^j \partial_j \pi_i
-M_0\Big] N\nonumber \\
&
-\frac{1}{2} C_S(N^\dagger N) (N^\dagger N)
-\frac{1}{2} C_T (N^\dagger \sigma_i N) (N^\dagger \sigma_i N)\, ,
\label{eq:L_0}
\end{align}
where $m_\pi$ is the pion mass,
$M_0$ is the bare nucleon mass,
$f_\pi=$ 92 MeV is the pion decay constant,
$g_A=1.26$ is the nucleon axial-vector coupling constant, $C_S$ and $C_T$ are low-energy constants, and $i = x, y, z$.
Throughout this work we adopt the convention that repeated indices imply the summation; additional conventions and notation details are reported in Appendix~\ref{app:conv}. At leading order, $m_\pi$ is the physical pion mass, whereas $M_0$ is regularization dependent and must be properly tuned for a given ultraviolet cutoff.

Establishing a rigorous
power-counting scheme in chiral effective field theory is currently
a subject of debate \cite{kap96,nog05,val06,epe13,son17}.
Our power 
counting gives an expansion in the number of pion field variables, 
in this work truncated at the quadratic level. This truncation then
defines 
the interaction between our degrees of freedom: pions and 
nucleons.
As in calculations
in which
nucleons are the only active
degrees
of freedom,
we solve the Schr\"odinger equation for the states of our
system
using this truncated interaction at all orders. Therefore, we consider
this to
be a leading-order calculation. The nucleon kinetic energy has been
promoted 
(as it is in other real-space interactions) since, with the nucleons
on a continuum,
the kinetic energy is required to have a well behaved Hamiltonian with
physical
states.
In principle, going to higher order is straightforward -- higher order
Lagrangians 
would include more pion interactions. Since pions are bosons, the
computational
complexity of the problem might increase, but we do not expect any
fundamental 
difficulties. If other power counting methods were to be devised, we
foresee
no major difficulty in modifying the techniques described here for
those possible future choices.

The conjugate momenta of the nucleon fields are defined as
\begin{align}
\Pi_N &=  \frac{\partial \mathcal{L}_0}{\partial(\partial_0 N)}=i N^\dagger,\nonumber\\
\Pi_{N^\dagger}&=\frac{\partial\mathcal{L}_0}{\partial(\partial_0 N^\dagger)}= 0,
\end{align}
while for the pions,
\begin{align}
\label{eq:conjfields}
\Pi_{k}&= \frac{\partial \mathcal{L}_0}{\partial(\partial_0 \pi_k)}=  \partial_0 \pi_k - \frac{1}{4f_\pi^2}\epsilon_{ijk} \pi_j N^\dagger  \tau_i   N\, .
\end{align}
The Hamiltonian can be written as a sum of three terms,
\[
H=H_{\pi\pi}+H_{\pi N}+H_{NN}.
\]
The pion Hamiltonian $H_{\pi\pi}$ is given by
\begin{equation}
\label{eq:Hpipi}
H_{\pi\pi}= \int d^3 x \frac{1}{2} \Big[ \Pi_{i}^2(\bvec{x}) +(\nabla \pi_i(\bvec{x}))^2 + m_\pi^2 \pi_i^2(\bvec{x})\Big]\, , 
\end{equation}
where the standard conventions adopted for the gradient are given in Appendix \ref{app:conv}.
The pion-nucleon interaction Hamiltonian reads
\begin{align}
\label{eq:HpiN}
H_{\pi N} &=\int d^3x \left[ \frac{g_A}{2f_\pi}  N^\dagger (\bvec{x}) \tau_i  \sigma^j \partial_j \pi_i(\bvec{x}) N(\bvec {x}) \right. \nonumber\\
&+\left. \frac{1}{4f_\pi^2}\epsilon_{ijk} \pi_j(\bvec{x}) \Pi_{k}(\bvec{x}) N^\dagger(\bvec{x})  \tau_i   N(\bvec{x})\right].
\end{align}
The first term is the axial-vector pion-nucleon coupling, and the second (referred to as the Weinberg-Tomozawa
term) is the contact interaction with two factors of the pion field interacting with the nucleon at a single point~\cite{Scherer:2009bt}.
Finally, the nucleon Hamiltonian is given by 
\begin{align}
\label{eq:HNN}
H_{NN} =& \int d^3x \Big[ N^\dagger (\bvec x)  \Big(-\frac{\nabla^2}{2M_0}  + M_0\Big) N(\bvec x) \nonumber\\
&+\frac{1}{2} C_S N^\dagger(\bvec x) N(\bvec x) N^\dagger(\bvec x) 
N(\bvec x)+ \nonumber \\
&  \frac{1}{2} C_T N^\dagger(\bvec x) \sigma_i N(\bvec x)  N^\dagger(\bvec x) \sigma_i N(\bvec x)\Big],
\end{align}
where $C_S$ and $C_T$ are two low-energy constants (LEC) that have to be fitted against two-nucleon properties.  

\subsection{Pion fields in the Schr\"odinger picture}
\label{sec:pion}
We work in the Schr\"odinger picture, where the pion fields and their conjugate momenta are time independent, and obey the canonical commutation relations, 
\begin{align}
[\pi_i(\bvec{x}),\pi_j(\bvec{y})]&=[\Pi_i(\bvec{x}),\Pi_j(\bvec{y})]=0,\nonumber\\
[\pi_i(\bvec{x}),\Pi_j(\bvec{y})]&=i\delta_{ij}\delta^{(3)}(\bvec{x}-\bvec{y}).
\label{eq:can_comm}
\end{align}
Let us perform a plane-wave expansion in a box of size $L$ with periodic boundary conditions, implying that the allowed momenta are discretized,
\begin{equation}
\bvec k=\frac{2\pi}{L} (n_x,n_y,n_z), \text{ with } n_i=0,\pm 1,\pm 2,\dots
\end{equation}
This discretization introduces an infrared cutoff on the three-momentum of the pions, proportional to the inverse of the size of the box. To avoid infinities, the theory is regularized introducing an ultraviolet cutoff for the three-momentum of the pions, such that $k\equiv |\bvec k|\leq k_c$. The Fourier expansions read
\begin{align}
\pi_i(\bvec{x}) &= \frac{1}{\sqrt{L^3}} \sum_{\bvec k} \pi_{i\bvec k} e^{i\bvec{k} \cdot \bvec x},\nonumber\\
\Pi_i(\bvec{x}) &= \frac{1}{\sqrt{L^3}} \sum_{\bvec k} \Pi_{i\bvec k} e^{i\bvec{k} \cdot \bvec x}\, .
\label{eq:fourier}
\end{align}
Since the fields are hermitian, the mode operators are such that $\pi^\dagger_{i\bvec{k}}=\pi_{i\bvec {-k}}$ and $\Pi^\dagger_{i\bvec {k}}=\Pi_{i\bvec {-k}}$.  The canonical commutation relations of Eq.~(\ref{eq:can_comm}) imply
\begin{align}
[\pi_{i \bvec{k}},\pi_{j \bvec{k}^\prime}]&=[\Pi_{i \bvec{k}},\Pi_{j \bvec{k}^\prime}]=0,\nonumber\\
[\pi_{i \bvec{k}},\Pi_{j \bvec{k}^\prime}]&=i\delta_{ij}\delta_{\bvec {k}-\bvec{k}^\prime}.
\label{eq:can_comm2}
\end{align}
When expressed in terms of the pion modes, the free pion Hamiltonian of Eq.~(\ref{eq:Hpipi}) describes a collection of harmonic oscillators with frequencies 
$\omega_{\bvec{k}}= \sqrt{k^2+m_\pi^2}$,
\begin{align}
\label{eq:Hpipi_k}
H_{\pi\pi}&= \sum_{\bvec k} \sum_i \left[ \frac{1}{2} \Pi_{i \bvec{k}}^2 +\frac{1}{2}\omega_{\bvec k}^2 \pi_{i \bvec{k}}^{2}\right].
\end{align}
The latter can be quantized by defining the creation and annihilation operators,
\begin{align}
a_{i\bvec k}&=\frac{1}{\sqrt{2\omega_{\bvec k}}}(\omega_{\bvec k}\pi_{i\bvec{k}}+i\Pi_{i\bvec{k}})\nonumber\\
a^\dagger_{i\bvec{k}}&=\frac{1}{\sqrt{2\omega_{\bvec k}}}(\omega_{\bvec k}\pi^\dagger_{i\bvec{k}}-i\Pi^\dagger_{i\bvec{k}})\, ,
\label{eq:cr_ann}
\end{align}
which are independent for each mode, and satisfy the canonical commutation relations,
\begin{equation}
 [a_{i\bvec{k}},a_{j\bvec{q}}^{\dagger}]=\delta_{ij}\delta_{\bvec{k}\bvec{q}}\, .
\end{equation}
Using Eq.~(\ref{eq:cr_ann}) to express $\pi_{i \bvec{k}}$ and $\Pi_{i \bvec{k}}$ in Eq.~(\ref{eq:fourier}) in terms of the creation and annihilation operators, we
recover the usual expansion for the pion field operator and its conjugate momentum, 
\begin{align}
\pi_i(\bvec{x}) &=
\frac{1}{\sqrt{2L^3}} \sum_{\bvec{k}} \frac{1}{\sqrt{\omega_{\bvec{k}}}}
\left [ a_{i\bvec{k}} e^{i\bvec{k} \cdot \bvec{x}}
+ a^\dagger_{i\bvec{k}} e^{-i\bvec{k} \cdot \bvec{x}}
\right ],\nonumber\\
\Pi_i(\bvec{x}) &=
\frac{-i}{\sqrt{2L^3}} \sum_{\bvec{k}} \sqrt{\omega_{\bvec{k}}}
\left [ a_{i\bvec{k}} e^{i\bvec{k} \cdot \bvec{x}}
- a^\dagger_{i\bvec{k}} e^{-i\bvec{k} \cdot \bvec{x}}
\right ]\, .
\end{align}

To implement this formalism in our quantum Monte Carlo
algorithm, it is convenient to rewrite the sum of Eq.~(\ref{eq:fourier}) in such a way that
$\bvec{k}$ is included and $-\bvec{k}$ is not. Specifically, if $k_z\neq 0$, then $k_z > 0$; if $k_z=0$ and $k_y\neq 0$, then $k_y > 0$; 
and if $k_z=k_y=0$, then $k_x \geq 0$. Let us define for $k \neq 0$,
\begin{align}
\pi_{i\bvec{k}}^c&=\frac{1}{\sqrt{2}}(\pi_{i\bvec{k}}+\pi_{i\bvec{-k}}),\nonumber\\
\pi_{i\bvec{k}}^s&=\frac{i}{\sqrt{2}}(\pi_{i\bvec{k}}-\pi_{i\bvec{-k}}),\, 
\label{eq:piksc}
\end{align}
while for $k = 0$ we have $\pi_{i0}^c=\pi_{i0}/\sqrt{2}$ and $\pi_{i 0}^s=0$. Employing analogous definitions for the conjugate momenta, the Fourier expansion of Eq.~(\ref{eq:fourier}) reads
\begin{align}
\pi_i(\bvec{x}) &= \sqrt{\frac{2}{L^3}} {\sum_{\bvec k}}^\prime[ \pi_{i\bvec{k}}^c \cos(\bvec{k} \cdot \bvec{x}) + \pi_{i\bvec{k}}^s \sin(\bvec{k} \cdot \bvec{x})], \nonumber\\
\Pi_i(\bvec{x}) &= \sqrt{\frac{2}{L^3}} {\sum_{\bvec k}}^\prime[ \Pi_{i\bvec{k}}^c \cos(\bvec{k} \cdot \bvec{x}) + \Pi_{i\bvec{k}}^s \sin(\bvec{k} \cdot \bvec{x})]\,,
\label{eq:fourierp}
\end{align}
where hereafter we adopt the convention of a primed sum to indicate that it is
over the set of $\bvec{k}$ described above. The commutation rules for $\pi_{i\bvec{k}}^{c,s}$ and $\Pi_{i\bvec{k}}^{c,s}$
follow from those of Eq.~(\ref{eq:can_comm2}). The only nonvanishing ones, valid also for $k=0$, are
\begin{align}
[\pi_{i \bvec{k}}^c,\Pi_{j \bvec{k}^\prime}^c]&=i\delta_{ij}\delta_{\bvec{k}\bvec{k}^\prime},\nonumber\\
[\pi_{i \bvec{k}}^s,\Pi_{j \bvec{k}^\prime}^s]&=i\delta_{ij}\delta_{\bvec{k}\bvec{k}^\prime},
\label{eq:can_comm3}
\end{align}
where we dropped the contribution proportional to $\delta_{\bvec{k}-\bvec{k}^\prime}$, as in the primed sums these cases are excluded.  

The pion Hamiltonian of Eq.~(\ref{eq:Hpipi_k}) becomes
\begin{eqnarray}
H_{\pi\pi} &=&
{\sum_{\bvec k}}^\prime\sum_i \left[ \frac{1}{2} \Pi^{c\,2}_{i\bvec k}
+ \frac{1}{2} \Pi^{s\,2}_{i\bvec k} \right .
\nonumber\\
&& \left. +\frac{1}{2}(k^2+m_{\pi}^2)
(\pi_{i \bvec{k}}^{c\,2}+\pi_{i \bvec{k}}^{s\,2})\right]\, .
\end{eqnarray}

In our simulations, in exact analogy to working in the
position operator eigenstates of the usual harmonic oscillator,
we work in the eigenbasis of the mode amplitude operators,
$\pi^{c,s}_{i\bvec k}$. 
Wave functions which are the overlaps of our states with this basis,
represent the states.
The momentum operators conjugate to
$\pi^{c,s}_{i\bvec k}$ are the generators of translations of these
amplitudes, and therefore when operating on a state represented
in this basis, they give the derivative of the wave function in the usual way,
\begin{equation}
\Pi_{i\bvec{k}}^{c,s} \to
-i \frac{\partial}{\partial \pi_{i\bvec{k}}^{c,s}}.
\end{equation}
Using the latter relation, the free pion Hamiltonian operating
on the state becomes the differential
operator
\begin{align}
\label{eq:Hpipi_k2}
H_{\pi\pi}&= {\sum_{\bvec k}}^\prime\sum_i \left[ - \frac{1}{2} \frac{\partial^2}{\partial \pi_{i\bvec{k}}^{c\,2}}  - \frac{1}{2} \frac{\partial^2}{\partial \pi_{i\bvec{k}}^{s\,2}} \right. \nonumber\\
&\left. +\frac{1}{2}(k^2+m_{\pi}^2) (\pi_{i \bvec{k}}^{c\,2}+\pi_{i \bvec{k}}^{s\,2})\right]\, ,
\end{align}
operating on the wave function.

The ground-state wave function for the pion modes is analogous to that describing the positions of a collection of quantum harmonic oscillators,
\begin{align}
\Psi_{0}(\boldsymbol{\pi}^{c,s})=\exp{\left[-{\sum_{\bvec k}}^\prime\sum_i \frac{\omega_k}{2}( \pi_{i \bvec{k}}^{c\,2}+\pi_{i \bvec{k}}^{s\,2})\right]}\, ,
\end{align}
where we use the symbol $\boldsymbol{\pi}^{c,s}$ to denote the full set of $ \pi_{i\bvec{k}}^{c}$ and $\pi_{i \bvec{k}}^{s}$.
When pion-nucleon interactions are accounted for in the Hamiltonian, the solution of the Schr\"odinger equation is no longer in closed form. We will employ QMC 
methods to tackle this problem when one and two nucleons are present in the system under study.

We define the Cartesian isospin field
\begin{equation}
\psi_i(\bvec{x}) \equiv \frac{1}{\sqrt{L^3}}
\sum_{\bvec{k}} a_{i\bvec{k}} e^{i\bvec{k} \cdot \bvec{x}},
\end{equation}
which obeys the commutation relations
\begin{equation}
\left [\psi_i(\bvec{x}),\psi^\dagger_j(\bvec{x}') \right ]
= \delta_{ij}
\frac{1}{L^3}\sum_{\bvec{k}} e^{i\bvec{k} \cdot (\bvec{x}-\bvec{x}^\prime)}.
\end{equation}
In the limit of infinite ultraviolet cutoff, the latter expression tends to $\delta^{(3)}(\bvec{x}-\bvec{x}^\prime)$, as prescribed
by the canonical commutation relations. For a finite cutoff $k_c$, the delta function will be smeared over
a volume proportional to $k_c^{-3}$.
The corresponding Cartesian isospin
component pion density operator is defined as
\begin{eqnarray}
\rho_i (\bvec{x}) =
\psi^\dagger_i(\bvec{x}) \psi_i(\bvec{x}).
\end{eqnarray}
We can also define density operators associated to different charge states
\begin{align}
\label{eq:dens_charge}
\rho_{\pi^+}(\bvec{x}) &=
\left [\frac{\psi^\dagger_x(\bvec{x})-i\psi^\dagger_y(\bvec{x})}{\sqrt{2}} \right ]
\left [\frac{\psi_x(\bvec{x})+i\psi_y(\bvec{x})}{\sqrt{2}} \right ],
\nonumber\\
\rho_{\pi^-}(\bvec{x}) &=
\left [\frac{\psi^\dagger_x(\bvec{x})+i\psi^\dagger_y(\bvec{x})}{\sqrt{2}} \right ]
\left [\frac{\psi_x(\bvec{x})-i\psi_y(\bvec{x})}{\sqrt{2}} \right ],
\nonumber\\
\rho_{\pi^0}(\bvec{x}) &=
\psi^\dagger_z(\bvec{x}) \psi_z(\bvec{x}) \,.
\end{align}
The densities are evaluated by using Eqs.~(\ref{eq:cr_ann}) and~(\ref{eq:piksc}) to transform the expressions above into 
a form suitable to be evaluated with wave functions containing the amplitudes of the pion modes.

Unlike standard Green's function
Monte Carlo (GFMC) calculations~\cite{car87,Carlson:2014vla}, the sum of the nucleon charges in our QMC simulations,
\begin{equation}
Q_N=\sum_{i=1}^A \frac{\left(1+\tau_z^i\right)}{2}\,,
\end{equation}
is not conserved configuration by configuration. This is due to the fact that the total charge of the system includes that of the charged pions, 
\begin{equation}
\label{eq:totalcharge}
Q=Q_N+Q_\pi,
\end{equation}
with $Q_\pi\equiv(N_+-N_-)$. The charged pion number operators are defined as
\begin{equation}
N_\pm=\sum_{\bvec{k}} a^\dagger_{\pm\bvec{k}}a^{\phantom{\dagger}}_{\pm\bvec{k}},
\end{equation}
where the creation and annihilation operators -- see Appendix~\ref{app:conv} for our conventions on the fields associated with charged pions -- are given by 
\begin{align}
a_{\pm\bvec{k}}^\dagger &=\frac{1}{\sqrt{2}}\left[ a_{x\bvec{k}}^\dagger \mp i a_{y\bvec{k}}^\dagger\right], \nonumber \\
a_{\pm\bvec{k}} &= \frac{1}{\sqrt{2}}\left[a_{x\bvec{k}} \pm i a_{y\bvec{k}}\right]\, .
\label{eq:charged_a}
\end{align}
while for the neutral pion, 
\begin{equation}
a_{0\bvec{k}}= a_{z\bvec{k}}\, .
\label{eq:neutral_a}
\end{equation}
The pion-charge is evaluated expressing the Cartesian isospin
creation and annihilation operators in terms of the modes of the pion field. 

It simplifies some expressions to combine the isospin
components into vectors in the usual way and define the pion mode
amplitudes and their conjugate momenta as the isospin vectors,
\begin{eqnarray}
\label{eq:pionvectors}
\bvec \Pi^{c,s}_{\bvec k} &=&
 \Pi^{c,s}_{x\bvec k}\bhat x
+\Pi^{c,s}_{y\bvec k}\bhat y
+\Pi^{c,s}_{z\bvec k}\bhat z,
\nonumber\\
\bvec \pi^{c,s}_{\bvec k} &=& 
\pi^{c,s}_{x\bvec k}\bhat x
+\pi^{c,s}_{y\bvec k}\bhat y
+\pi^{c,s}_{z\bvec k}\bhat z.
\end{eqnarray}
The pion charge operator becomes
\begin{eqnarray}
Q_\pi &=& -\bhat z \cdot {\sum_{\bvec k}}' \left [
\bvec \pi^c_{\bvec k} \times \bvec \Pi^c_{\bvec k}
+\bvec \pi^s_{\bvec k} \times \bvec \Pi^s_{\bvec k} \right ],
\end{eqnarray}
or as a differential operator on a wave function,
\begin{align}
Q_\pi&=i{\sum_{\bvec k}}^\prime\left[\pi_{x\bvec{k}}^c\frac{\partial}{\partial \pi_{y\bvec{k}}^c}
-\pi_{y\bvec{k}}^c\frac{\partial}{\partial \pi_{x\bvec{k}}^c}\right. \nonumber\\ 
&+\left . \pi_{x\bvec{k}}^s\frac{\partial}{\partial \pi_{y\bvec{k}}^s}
-\pi_{y\bvec{k}}^s\frac{\partial}{\partial \pi_{x\bvec{k}}^s}
\right].
\end{align}

\section{Quantum Monte Carlo}
\label{sec:qmc0}
With our periodic box and the pion momentum cutoff, we now have a finite
number of degrees of freedom, and can now use real-space quantum Monte
Carlo methods to solve for the ground and low lying excited state
properties of $A$ nucleons. 
Our goal here is to be able to adapt variational Monte Carlo (VMC),
GFMC, and Auxiliary field diffusion Monte Carlo (AFDMC) \cite{sch99} 
methods to include the pion degrees of freedom. We therefore need
to write our Hamiltonian in the $A$ nucleon sector along with the
pion fields, find good initial variational trial wave functions, and
describe how we include the additional terms in the propagators.
Note that, at variance with nuclear lattice approaches~\cite{lee09}, we adopt
a continuum representation for the eigenstates of the position operator.

\subsection{The quantum Monte Carlo Hamiltonian}
We write the pion operators using Eq.~(\ref{eq:pionvectors})
and the momentum operator conjugate to the particle position operator,
$\bvec r_i$, as $\bvec P_i$.
Since the number of nucleons is conserved, the 
Hamiltonian for the sector with $A$ nucleons
and the pion field can be written down immediately,
\begin{eqnarray}
\label{eq:hqmc}
H&=&H_N+H_{\pi\pi}+H_{AV}+H_{WT},
\nonumber\\
H_N &=& \sum_{i=1}^A \left [ \frac{P_i^2}{2M_P}+M_P
+\beta_K P_i^2 + \delta M\right ]
\nonumber\\
&&
+\sum_{i<j}^A \delta_{k_c}(\bvec r_i-\bvec r_j)
[C_S+C_T\bvec \sigma_i\cdot \bvec \sigma_j ],
\nonumber\\
H_{\pi\pi} &=& \frac{1}{2}{\sum_{\bvec k}}' \left [
|\bvec \Pi^c_{\bvec k}|^2+
\omega_{\bvec k}^2
|\bvec \pi^c_{\bvec k}|^2
+|\bvec \Pi^s_{\bvec k}|^2+
\omega_{\bvec k}^2
|\bvec \pi^s_{\bvec k}|^2
\right ],
\nonumber\\
H_{AV} &=&
\sum_{i=1}^A
\frac{g_A}{2 f_\pi} \sqrt{\frac{2}{L^3}}
{\sum_{\bvec k}}^\prime \left\{\bvec \sigma_i \cdot \bvec k
\left[\bvec \tau_i \cdot \bvec \pi_{\bvec k}^s \cos(\bvec k \cdot \bvec r_i)
\right. \right. \nonumber\\
&-& \left. \left. \bvec \tau_i\cdot \bvec \pi_{\bvec{k}}^c
\sin(\bvec k \cdot \bvec r_i)\right]\right\},
\nonumber\\
H_{WT} &=&\sum_{i=1}^A
\frac{1}{2f_\pi^2 L^3} \bvec \tau_i
\cdot \left [
\rule[-1.1em]{0em}{0em}
\right .
\nonumber\\
&&
{\sum_{\bvec k}}' \cos(\bvec k \cdot \bvec r_i) \bvec \pi^c_{\bvec k}
\times
{\sum_{\bvec q}}' \cos(\bvec q \cdot \bvec r_i) \bvec \Pi^c_{\bvec q}
\nonumber\\
&+&
{\sum_{\bvec k}}' \cos(\bvec k \cdot \bvec r_i) \bvec \pi^c_{\bvec k}
\times
{\sum_{\bvec q}}' \sin(\bvec q \cdot \bvec r_i) \bvec \Pi^s_{\bvec q}
\nonumber\\
&+&
{\sum_{\bvec k}}' \sin(\bvec k \cdot \bvec r_i) \bvec \pi^s_{\bvec k}
\times
{\sum_{\bvec q}}' \cos(\bvec q \cdot \bvec r_i) \bvec \Pi^c_{\bvec q}
\nonumber\\
&+& \left .
{\sum_{\bvec k}}' \sin(\bvec k \cdot \bvec r_i) \bvec \pi^s_{\bvec k}
\times
{\sum_{\bvec q}}' \sin(\bvec q \cdot \bvec r_i) \bvec \Pi^s_{\bvec q}
\right ],
\end{eqnarray}
where the sums over $i$ and $j$ are over the nucleons, $M_P$ is the physical
nucleon mass, and
$\delta_{k_c}(\bvec r_i-\bvec r_j)$ is a smeared out delta function
for the contact term, which we take to be
consistent with the cutoff employed for the pion modes,
\begin{equation}
\label{eq:deltakc}
\delta_{k_c}(\bvec r)=\frac{1}{L^3}\left(
1+2{\sum_{\bvec{k}}}'\cos(\bvec{k}\cdot\bvec{r})
\right).
\end{equation}

Although we report the low-energy constants for this choice
of the smeared out delta function in Sec.~\ref{sec:2nuc}, we also considered
a different functional form,
commonly used in local chiral-EFT potentials~\cite{gez14};
see Appendix~\ref{app:contact}.
There are, of course, many other possible choices for the
functional form of the smeared out delta function.
We are aware that there might be shortcomings in employing
Eq.~(\ref{eq:deltakc}) \cite{lov12,hut17}, and the
naive dimensional analysis power counting
that is behind them \cite{kap96,nog05}.
However, they are mitigated by the fact we focus on deuteron 
properties, which has a relatively small $d$-wave component, and 
we employ fixed pion masses. This regulator choice will be 
thoroughly analyzed, and might be revisited, for larger nuclei.
The low-energy constants $C_S$ and $C_T$ need to be adjusted to
reproduce two-nucleon observables.
We fit them using the deuteron and two neutrons in Sec.~\ref{sec:2nuc}.


Notice that we have two distinct mass counter terms in $H_{N}$. We call
$\beta_K$ the kinetic mass counter term and $\delta M$ the rest mass
counter term. The values are not simply related because we are employing a
cutoff on the three-momentum of the pion modes that explicitly breaks
Lorentz invariance. The kinetic energy bare mass is given by
$\frac{M_P}{1+2\beta_K M_P}$, while the bare rest mass is $M_P+\delta M$.

Our resulting field theory Hamiltonian is in the same form as the
Hamiltonian of a
nonrelativistic many-body quantum system, and all standard methods for
such a system can be applied.

In this work we will sometimes neglect the Weinberg-Tomozawa $H_{WT}$
term in our initial QMC calculations. In general we have found
that it is small enough to be included perturbatively.
This term is known to be relevant only in the isovector channel, and
the $s$-wave $\pi$N scattering length is relatively small \cite{rob78,wei92}.
These assumptions have to be carefully checked when
studying
$A\geqslant 3$ systems, and most likely will not hold for $p$-shell and larger nuclei.

\subsection{Trial wave functions}

Analogously to standard real-space QMC methods, we first construct
an accurate ground state trial wave function for the Hamiltonian.
In GFMC or AFDMC methods, the trial function performs the dual
role of lowering the statistical errors and constraining the
path integral to control the fermion sign or phase problem.
For small numbers of nucleons where the fermion sign/phase
problem is under control, our QMC methods will give exact
results within statistical errors independent of the trial function.
A good trial function in that case keeps the statistical errors small.

Standard GFMC and AFDMC methods use the position eigenbasis for the
nucleons. Here we add to this nucleon basis
the eigenbasis of the pion mode
amplitudes, and write our trial wave functions to be 
\begin{equation}
\label{eq:psitrspi}
\Psi_T(R,S,\Pi) = \langle RS\Pi|\Psi_T\rangle,
\end{equation}
where $R$ represents the $3A$ coordinates of the nucleons, $\Pi$ represents
the $3N_{\bvec k}$ pion mode amplitudes, and $S$ the spin-isospin of
the nucleons. 

If we assume the pion motion is significantly faster than the nucleons,
then a Born-Oppenheimer approximation where we initially neglect the nucleon
mass can guide our construction of a trial wave function for the full
dynamical system.
We therefore initially analyze the problem without
the nucleon kinetic energy
and the Weinberg-Tomozawa terms in the
Hamiltonian, assuming that they are smaller than the axial-vector pion-nucleon
terms. Defining
\begin{align}
\bvec B^c_{\bvec{k}}&\equiv\sqrt{\frac{2}{L^3}}
\frac{g_A}{f_\pi} \sum_{i=1}^A \bvec \tau_i \sin(\bvec k \cdot \bvec r_i)
\bvec \sigma_i \cdot \bvec k,\nonumber\\
\bvec B^s_{\bvec{k}}&\equiv -\sqrt{\frac{2}{L^3}} \frac{g_A}{f_\pi}
\sum_{i=1}^A \bvec \tau_i\cos(\bvec k \cdot \bvec r_i) 
\bvec \sigma_i \cdot \bvec k,
\end{align}
allows us to complete the squares in these terms of the Hamiltonian, yielding
\begin{eqnarray}
\label{eq:hfixed}
H_{\pi\pi} +H_{AV} &=& \frac{1}{2}{\sum_{\bvec k}}' \left [
|\bvec \Pi^c_{\bvec k}|^2+
\omega_{\bvec k}^2
|\bvec {\tilde \pi}^c_{\bvec k}|^2
+
|\bvec \Pi^s_{\bvec k}|^2+
\omega_{\bvec k}^2
|\bvec {\tilde \pi}^s_{\bvec k}|^2
\right .
\nonumber\\
&& \left .
 -\frac{1}{4\omega_k^2} \left (
\left | \bvec B^c_{\bvec{k}}\right |^2
+\left | \bvec B^s_{\bvec{k}}\right |^2 \right )
\right ]
\end{eqnarray}
with
$\bvec {\tilde\pi}_{\bvec{k}}^{c,s} \equiv \bvec \pi_{i\bvec{k}}^{c,s}
- \bvec B^{c,s}_{\bvec{k}}/2\omega_k^2$.

The $\bvec {\tilde\pi}_{\bvec k}^{c,s}$ operators do not commute
because of the nucleon spin-isospin operators contained in
$\bvec B_{\bvec k}^{c,s}$. 
If instead these spin-isospin operators were c-numbers, we could immediately
write the ground-state wave function for the pions. This suggests
taking the form for trial wave function to be
\begin{equation}
\langle RS\Pi|\Psi_T\rangle =
\langle RS\Pi|
\exp{\left[-{\sum_{\bvec k}}^\prime \frac{\omega_k}{2}(
|\bvec{\tilde\pi}_{\bvec{k}}^c|^2
+|\bvec {\tilde\pi}_{\bvec{k}}^s|^2)\right]}|\Phi\rangle\, ,
\end{equation}
where $\ket{\Phi}$ is an $A$ nucleon model state.
Writing
in terms of the original pion coordinates, this wave function becomes
\begin{align}
\label{eq:psitqmc}
&\langle RS\Pi|\Psi_T\rangle =
\langle RS\Pi|
\exp\left\{
-{\sum_{\bvec k}}^\prime \left [
\rule[-1.6em]{0em}{0em}
\frac{\omega_k}{2}(
|\bvec{\pi}_{\bvec{k}}^c|^2
+|\bvec {\pi}_{\bvec{k}}^s|^2)\right .
\right .
\nonumber\\
&
-\frac{\alpha_k}{2\omega_k}
\left (
\bvec \pi_{\bvec k}^c \cdot \bvec B^c_{\bvec k}
+\bvec \pi_{\bvec k}^s \cdot \bvec B^s_{\bvec k}
\right )
\nonumber\\
&
\left .
\left .
+\frac{1}{4}\omega_k\alpha_k^2 G_k^2 \sum_{i<j}^A
\bvec \tau_i\cdot\bvec \tau_j
\bvec\sigma_i\cdot \bvec k
\bvec\sigma_j\cdot \bvec k
\cos(\bvec k\cdot \bvec r_{ij})
\right] \right \}
|\Phi\rangle\, ,
\end{align}
where $\bvec r_{ij} = \bvec r_i-\bvec r_j$,
\begin{equation}
G_k=\frac{1}{\omega_k^2}\frac{g_A}{f_\pi}\sqrt{\frac{2}{L^3}}\,,
\end{equation}
and we drop terms that only contribute to the overall normalization.
We have also introduced the variational parameters $\alpha_k$, which
rescale the coupling for different momenta. 

Equation~(\ref{eq:psitqmc})
is the standard form we will take for our trial functions.
The two-body terms do not contain pion amplitudes; they look like
two-body correlations typically included
in variational calculations, and therefore they
could be replaced or modified with other correlation forms that
may be more convenient for calculations. The pion-nucleon correlation
terms look very much like the AFDMC propagators, as it would be expected
from the fact that the auxiliary fields in AFDMC can be thought of
as replacing the real pion fields.

\subsection{Nucleon model states}
\label{sec:two}
To complete our trial wave functions, we need to construct good trial
nuclear model states, $|\Phi\rangle$ of Eq.~(\ref{eq:psitqmc}).
We again are guided by previous experience with GFMC and AFDMC calculations.
The trial functions there are typically built from operator
correlated linear combinations of antisymmetric products of
single-particle orbitals. For example, in nuclear matter the trial function
is a Jastrow product of pair-wise operator correlations operating on
a Slater determinant of orbitals. Here we will begin by assuming that
the pion-nucleon correlations and the associated terms in
Eq.~(\ref{eq:psitqmc}) will include the long-range correlations.

Initially, we build our nuclear model state in the same way. However,
we include only short-range operator correlations; the remaining
terms in Eq.~(\ref{eq:psitqmc}) will include long-range correlations.
For the calculations described here, we need to construct model states
for one- and two-nucleon systems.
Since a single nucleon only interacts with the pion field, its model
state $|\Phi\rangle$ in Eq.~(\ref{eq:psitqmc})
is a spin-isospin state, i.e., proton up, proton down, neutron up, neutron
down, with no spatial dependence.

Two nucleons in our Hamiltonian interact via pion exchange and from the
short-range smeared-out contact interactions. A reasonably
good trial wave function for s-shell nuclei that contains
the major correlations can be constructed with a Jastrow operator
product multiplying an antisymmetric product of spin-isospin states.
The Jastrow factors go to zero exponentially to properly match
the separation energy of one nucleon. At short range, the Jastrow
factors solve the two-body Schr\"odinger equation. 

Local chiral-EFT interactions~\cite{tew13,gez13,gez14,lyn17}
at LO are usually written 
in the form
\begin{align}
V_{NN}(\bvec r_{ij})&=\sum_{p=1}^6 v^p(r_{ij}) O^p_{ij}\, ,
\label{eq:v(r)}
\end{align}
where the radial functions $v^p(r_{ij})$ are fully specified in Ref.~\cite{gez14}. 
The six operators $O^p_{ij}$ are  $1$, $\bvec \sigma_i \cdot \bvec \sigma_j$,
$S_{ij} =  3 (\bvec \sigma_i \cdot \bhat r_{ij})
(\bvec \sigma_j \cdot \bhat r_{ij}) -\bvec \sigma_i\cdot \bvec \sigma_j$
and each of those multiplied by $\bvec \tau_i\cdot\bvec \tau_j$.
We solve the two-body Schr\"odinger equation for the Jastrow factors
with this form of the interaction.

To fit the low-energy constants, we calculate in the two attractive
channels, the deuteron channel 
with total isospin $T = 0$, total spin $S=1$,
total angular momentum-parity 
$J^\pi=1^+$, with
deuteron binding energy $E=2.225$ MeV,
and the $S=0$, $T=1$, s-wave neutron-neutron channel.

The deuteron state can be written as
\begin{equation}
\label{eq:deuteron}
\ket{\Phi}^d= 
\left ( f_0^d(r_{12})+\frac{1}{\sqrt{8}} f_2^d(r_{12}) S_{12} \right ) | 0 \rangle^d ,
\end{equation}
where $f_l^d$ are radial functions, and $|0\rangle^d$ is a spin triplet,
isospin singlet state.
The deuteron Hamiltonian is
\begin{equation}
H^d = -\frac{\nabla^2_{r_{12}}}{2\mu}  + \tilde v_c^d(r_{12}) + \tilde v_t^d(r_{12}) S_{12},
\end{equation}
where $\mu$ is the reduced mass and
\begin{align}
\tilde v_c^d(r_{12}) &= v^1(r_{12})+v^3(r_{12})-3v^{4}(r_{12}),\nonumber\\
\tilde v_t^d(r_{12}) &= v^{5}(r_{12})-3v^{6}(r_{12}),
\end{align}
which leads to the
Schr\"odinger equation
for $F_l^d(r_{12}) \equiv r_{12} f_l^d(r_{12})$,
\begin{align}
-\frac{ {F_0^d}''}{2\mu} &+ \tilde v_c^d F_0^d + \sqrt 8 \tilde v_t^d F_2^d =E F_0^d, \nonumber\\
-\frac{{F_2^d}''}{2\mu} &+\frac{3 F_2^d}{\mu r_{12}^2} + (\tilde v_c^d -2\tilde v_t^d) F_2^d
+ \sqrt{8}  \tilde v_t^d F_0^d = E F_2^d,
\label{eq:coupled_d}
\end{align}
which can be readily integrated.

For the two-neutron case we have a $T=1$, $S=0$ spin-isospin state for
$|0\rangle^{nn}$,
\begin{equation}
\tilde{v}_c^{nn}(r_{12})=v^1(r_{12})+v^2(r_{12})-3v^3(r_{12})-3v^4(r_{12}),
\end{equation}
and we solve
\begin{align}
-\frac{{F_0^n}''}{2\mu} + \tilde v_c^{nn} F_0^{nn}=E F_0^{nn}\,,
\label{eq:coupled_nn}
\end{align}
where $F_l^{nn}(r_{12}) \equiv r_{12} f_l^{nn}(r_{12})$.

For the short-range interaction here, we use
\begin{align}
H_{NN}^{2N} &= \sum_{i=1}^2 \left [ \frac{P_i^2}{2M_P}+M_P
+\beta_K P_i^2 + \delta M\right ]\nonumber\\
&+ C_S \delta_{k_c}(\bvec{r}_{12})
+ C_T \delta_{k_c}(\bvec{r}_{12})\sigma_{12},
\label{eq:h_nn2n}
\end{align}
and we take $C_S$ and $C_T$ to be tunable constants. 

The wave function for the deuteron and for two neutrons is
Eq.~(\ref{eq:psitqmc})
using the
model state $\ket{\Phi}$ given by $\ket{\psi}^{d,nn}$.
When solving the corresponding differential Eqs. \eqref{eq:coupled_d} and \eqref{eq:coupled_nn}, we only
retain the contact contributions of the leading-order local chiral potential of Ref.~\cite{gez14}. The correlations arising from the one-pion exchange
term are dynamically generated when summing over the pion modes.

\subsection{Quantum Monte Carlo implementation}
\label{sec:qmc}
The variational Monte Carlo (VMC) algorithm exploits the stochastic Metropolis algorithm to evaluate the expectation value of a given many-body operator 
using a suitably parametrized trial wave function,
$\Psi_T(\bvec {\cal R})$, where we write $\bvec {\cal R}$ as an
abbreviation for $RS\Pi$ of Eq.~(\ref{eq:psitrspi}).
An integration over an $\bvec {\cal R}$ variable
stands for integration over the nuclear coordinates and pion field
amplitudes and a summation over the spin-isospin states.

The variational parameters of the trial wave function are found minimizing the the expectation value of the 
Hamiltonian,
\begin{equation}
E_V = \frac{\bra{\Psi_T}H\ket{\Psi_T}}{\product{\Psi_T}{\Psi_T}}.
\end{equation}
To this aim, in this work we employed the {\it linear method}, introduced in Ref.~\cite{tou07} that consists of diagonalizing a nonsymmetric
estimator of the Hamiltonian matrix in the basis of the wave function and its derivatives with respect to the parameters.
We have also adopted the heuristic procedure of Ref.~\cite{con17}, which suppresses instabilities that arise from the nonlinear dependence of the wave function on the variational parameters. 

The Green's function Monte Carlo method projects out of a trial wave function the lowest eigenstate $\ket{\Psi_0}$ of the Hamiltonian $H$ with nonzero overlap with $\ket{\Psi_T}$,
\begin{equation}
\ket{\Psi_0}\propto \lim_{\tau\to \infty} \exp\left[-(H-E_T)\tau\right]\ket{\Psi_T},
\end{equation}
where $E_T$ controls the normalization. In most cases the propagator
$\exp\left[-(H-E_T)\tau\right]$ cannot be calculated analytically. A repeated application of a short-time propagator can instead be used. This can be shown by inserting a sequence of completeness relations between each short-time propagator,
\begin{align}
\label{eq:prop_short}
&\Psi_0(\bvec {\cal R})\equiv \product{\bvec {\cal R}_N}{\Psi_0}=
\int d\bvec {\cal R}_1 \cdots d\bvec{\cal R}_{N-1}
\nonumber\\
&\left(\prod_{i=0}^{N-1} \bra{\bvec{\cal R}_{i+1}}\exp\left[-(H-E_T)\delta\tau\right]\ket{\bvec{\cal R}_i}\right) \product{\bvec {\cal R}_0}{\Psi_T}\, .
\end{align}
Monte Carlo techniques are used to sample the $\bvec {\cal R}_i$ in the
propagation at each imaginary time-step.
For a detailed description of the algorithm, the reader is referred to the review of Ref.~\cite{Carlson:2014vla} and references therein.

Both the one- and two-nucleon Hamiltonians are a sum of operators that in general do
not commute.
The short-time propagators of Eq.~(\ref{eq:prop_short}),
\begin{equation}
G(\bvec {\cal R}',\bvec {\cal R})=\bra{\bvec {\cal R}'}  \exp\left[-(H-E_T)\delta\tau\right] \ket{\bvec {\cal R}},
\end{equation}
can be split into kinetic and potential pieces using
the Trotter breakup formula.
Since we consider two versions of the Hamiltonian (including and omitting the Weinberg-Tomozawa term), we need to consider two distinct versions of the propagator,
\begin{align}
\label{eq:axial}
&G_{\rm AV}(\bvec {\cal R}',\bvec {\cal R})= \exp\left[\delta \tau E_T \right] \bra{\bvec {\cal R}'} \exp\left[-\delta \tau H_{AV}/2 \right]\nonumber\\
&\times \exp\left[-\delta \tau H_{\pi\pi} \right] \exp\left[-\delta \tau T \right] \exp[-\delta \tau V_{NN}]\nonumber\\
&\exp\left[-\delta \tau H_{AV} /2 \right]\ket{\bvec {\cal R}}
\end{align}
and
\begin{align}
\label{eq:full}
&G_{\rm WT}(\bvec {\cal R}',\bvec {\cal R})=\exp\left[\delta \tau E_T \right] \bra{\bvec {\cal R}'} \left(1-\delta \tau H_{WT}\right)\nonumber\\
&\times  \exp\left[-\delta \tau H_{AV}/2 \right] \exp\left[-\delta \tau H_{\pi\pi} \right]\exp\left[-\delta \tau T \right]\nonumber\\
&\times \exp[-\delta \tau V_{NN}]\exp\left[-\delta \tau H_{AV}/2 \right] \ket{\bvec {\cal R}},
\end{align}
with $V_{NN}=0$ in the one-nucleon case, and $V_{NN}=C_S \delta_{k_c}(\bvec{r}_{12})+ C_T \delta_{k_c}(\bvec{r}_{12})\sigma_{12}$ for the two-nucleon system.

The Euclidean time propagator associated with the nonrelativistic kinetic energy of the nucleons $T$ gives rise to a free diffusion process described by the propagator:
\begin{align}
\label{eq:prop_kin}
G_T(\bvec {\cal R}',\bvec {\cal R})&=\bra{\bvec {\cal R}'}\exp\left[-T\delta\tau\right]\ket{\bvec {\cal R}} \nonumber\\
&=\left[\frac{1}{\lambda^3 \pi^{3/2}}\right]^A \exp\left[ -\frac{(\bvec {\cal R}-\bvec {\cal R}')^2}{\lambda^2} \right],
\end{align}
with $\lambda=\sqrt{2\delta\tau/M}$.

The free propagator of the pion modes is the product of one-dimensional harmonic-oscillator Green's functions,
\begin{align}
&G_{\pi\pi}(\bvec {\cal R}',\bvec {\cal R})=\bra{\bvec {\cal R}'}\exp\left[-H_{\pi\pi}\delta\tau\right]\ket{\bvec {\cal R}} \nonumber\\
&\quad ={\prod_{\bvec{k}}}^\prime \prod_i G_{\rm HO}({\pi_{i\bvec{k}}^{c\, \prime}},\pi_{i\bvec{k}}^c)G_{\rm HO}({\pi_{i\bvec{k}}^{s\,\prime}},\pi_{i\bvec{k}}^s),
\end{align}
where
\begin{align}
&G_{\rm HO}(\pi_{i\bvec{k}}^{c,s\, \prime},\pi_{i\bvec{k}}^{c,s})=\left(\frac{\omega_k}{2\pi \sinh(\omega_k \delta\tau)}\right)^{1/2} \nonumber\\
&\exp\left[-\frac{\omega_k}{2\sinh(\omega_k\delta\tau)}\left[ (\pi_{i\bvec{k}}^{c,s\, \prime\, 2}+\pi_{i\bvec{k}}^{c,s\, 2})\cosh(\omega_k\delta\tau) \right. \right. \nonumber\\
&\left. -2 \pi_{i\bvec{k}}^{c,s\, \prime}\pi_{i\bvec{k}}^{c,s} \right] \bigg]\, .
\end{align}
The importance sampled version of this Green's function is \cite{kal97}
\begin{align}
\tilde{G}_{\rm HO}(\pi_{i\bvec{k}}^{c,s\, \prime},\pi_{i\bvec{k}}^{c,s})&=\frac{\phi_0(\pi_{i\bvec{k}}^{c,s\, \prime})}{\phi_0(\pi_{i\bvec{k}}^{c,s})}
\exp\left[\frac{\omega_k\delta\tau}{2}\right]\nonumber\\
&\times G_{\rm HO}(\pi_{i\bvec{k}}^{c,s\, \prime},\pi_{i\bvec{k}}^{c,s})\, ,
\end{align}
where $\phi_0$ is the ground-state wave function of the harmonic oscillator with energy $\omega_k/2$. The above equation can be cast in the form
\begin{align}
\label{eq:HOgreens}
\tilde{G}_{\rm HO}(\pi_{i\bvec{k}}^{c,s\, \prime},\pi_{i\bvec{k}}^{c,s})= \left( \frac{\omega_k}{\pi(1-e^{-2\omega_k\delta\tau})} \right)^{1/2}
\nonumber\\
\times \exp\left[-\frac{\omega_k(\pi_{i\bvec{k}}^{c,s\, \prime}-e^{-\omega\delta\tau}\pi_{i\bvec{k}}^{c,s})^2}{1-e^{-2\omega_k\delta\tau}} \right],
\end{align}
which is a Gaussian centered at $e^{-\omega_k\delta\tau}\pi_{i\bvec{k}}^{c,s}$ with variance $(1-e^{-2\omega_k\delta\tau})/(2\omega_k)$. Notice that for $\delta\tau\to\infty$ we have $\tilde{G}_{\rm HO}(\pi_{i\bvec{k}}^{c,s\, \prime},\pi_{i\bvec{k}}^{c,s}) \to \phi_0^2(\pi_{i\bvec{k}}^{c,s\, \prime})$, and for $\delta\tau\to 0$ we recover the free-particle propagator.
The propagator of Eq.~(\ref{eq:full}) contains pion derivatives.
As a first-order approximation, we act with the pion derivatives present in $H_{WT}$ on the propagator for the harmonic oscillators $\tilde{G}_{\rm HO}$. This procedure omits possible terms rising from the commutators, and it is analogous to the one used to implement spin-orbit propagator used in other quantum Monte Carlo methods for many-nucleon systems \cite{sar03}.

The direct calculation of the expectation value of an operator $O$ other than the Hamiltonian, and that does not commute with $H$, corresponds to a matrix element which is usually called a ``mixed estimator,"
\begin{eqnarray}
\expected{O}_m = \frac{\bra{\Psi_T}O\ket{\Psi_0}}{\product{\Psi_T}{\Psi_0}}.
\end{eqnarray}
The ``extrapolation method,'' which combines results of diffusion and variational simulations, is one of the most used to compute mixed estimators. However, its accuracy relies on the
quality of the trial wave function and, even in the case of accurate trial wave functions, the bias of the extrapolated estimator is difficult to assess. For these reasons, we use the ``forward walking method,'' discussed in detail in Ref.~\cite{cas95}, to evaluate the pion density. This method relies on the calculation of the asymptotic offspring of walkers coming from the branching term to compute
the exact estimator,
\begin{equation}
\label{eq:exact_est}
\expected{O}_e = \frac{\bra{\Psi_0}O\ket{\Psi_0}}{\product{\Psi_0}{\Psi_0}}.
\end{equation}

\section{Results}
\label{sec:results}
All the results are obtained considering a cell in momentum space, in which the sums over the $\bvec{k}$
wave vectors are limited by the spherical cutoff $\omega^s_c=\sqrt{k_c^2+m_\pi^2}$, where
\begin{eqnarray}
\frac{4\pi k_c^3}{3} = \left (\frac{2\pi}{L}\right )^3 N_{\bvec k},
\end{eqnarray}
$N_{\bvec k}$ being the number of $\bvec k$ vectors in the unprimed sums.
The number of wave vectors, in the primed sums, in each of the first 10 shells is (1,3,6,4,3,12,12,6,15,12,12). When $\bvec k$ has no zero components
there are 6 pion coordinates associated with each $\bvec{k}$, corresponding to the sine and cosine components of the three Cartesian
isospin coordinates. We set the pion mass to the average of the masses of the neutral and charged pions, $m_\pi=(m_{\pi_0}+2m_{\pi_\pm})/3=138.04$ MeV.
For the nucleon physical mass we used $M_P=2\mu=938.92$ MeV,
where $\mu$ is the reduced mass
given by
$1/\mu=1/M_{\rm proton}+1/M_{\rm neutron}$.

It is worth noting that for one nucleon there is no node-crossing, because no fermion exchange occurs with only one fermion.
For the two nucleon case the node-crossing is also zero. We expect s-shell nuclei to have a mild fermion sign or phase problem, as occurs in potential models.
Whenever energies are computed, both the full propagator and the propagator omitting the Weinberg-Tomozawa term are used. For all other estimators we have limited ourselves to the latter case only.

\subsection{Mass renormalization}
\label{sec:energy}
Since our choice for the momentum cutoff is not Lorentz invariant, the
two mass counter terms appearing in the Hamiltonians, $\beta_K$ and $\delta M$,
are not simply related.  The kinetic mass counter term coefficient $\beta_K$
is determined by requiring that
the nucleon diffuses with the physical mass $M_P=938.92$ MeV for long
imaginary-times, and $\delta M$ is set so that the ground state energy of
one nucleon is also the physical mass $M_P$.

To determine $\beta_K$, let us consider the diffusion of (classical) particles that
are initially at the origin, $C(\bvec{r}=0,\tau=0)=\delta^{(3)}(\bvec{r})$. The solution for the diffusion equation
\begin{eqnarray}
\frac{\partial C(\bvec{r},\tau)}{\partial \tau} = \frac{\nabla^2}{2M_K} C(\bvec{r},\tau)
\end{eqnarray}
is a Gaussian centered at the origin and with variance $\tau/M_K$. Multiplying the diffusion equation by $\bvec{r}^2$ and integrating over $\bvec{r}$, we get for the mean square displacement $\expected{r^2(\tau)}=3\tau/M_K$ and the kinetic mass of the nucleon can be computed from the slope of $\expected{r^2(\tau)}$. 

\begin{figure}[!htb]
\centering
\includegraphics[angle=-90,width=\linewidth]{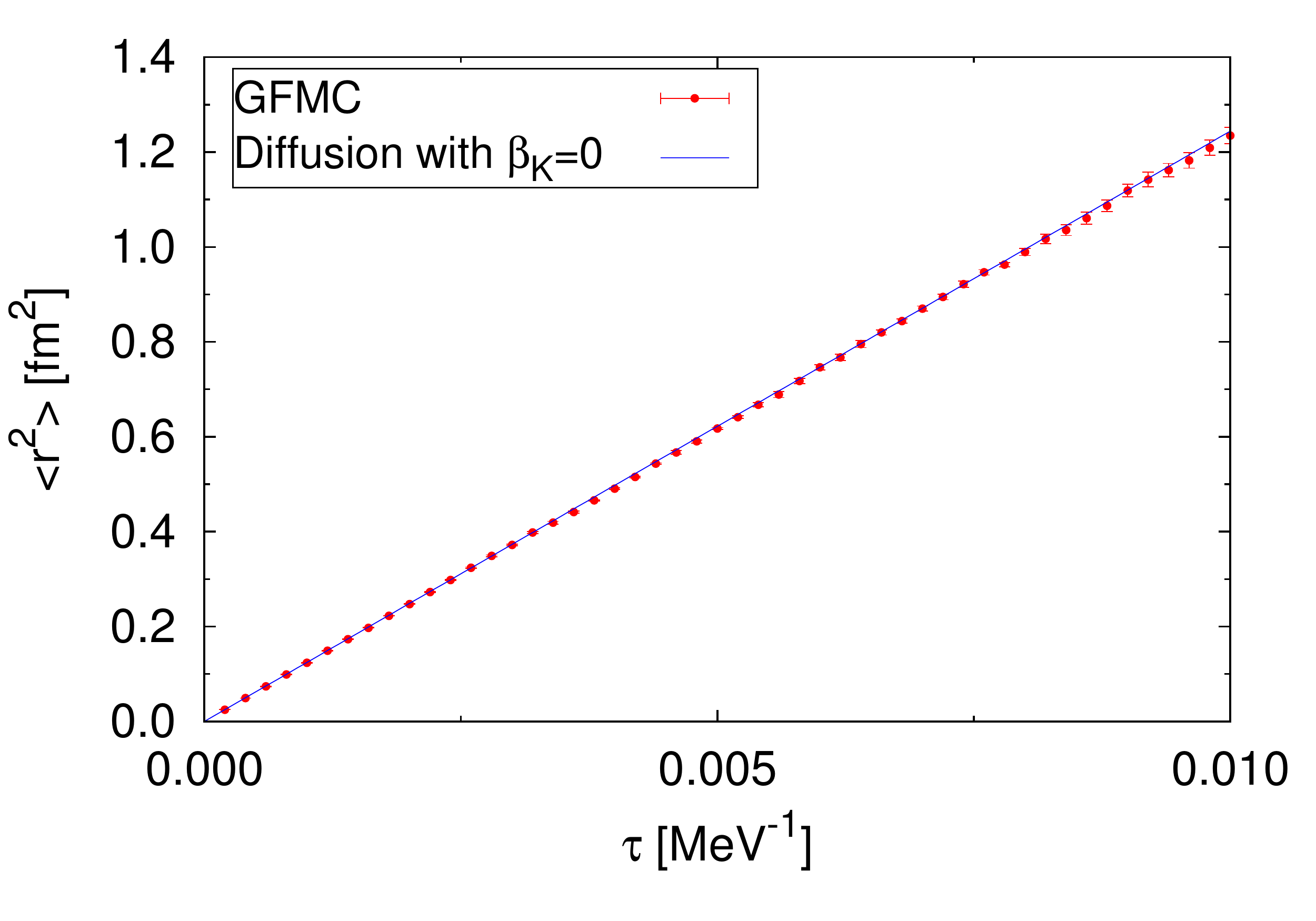}
\caption{(Color online)
Mean-square displacement $\expected{r^2}$ as a function of the imaginary-time $\tau$. The (blue) curve stands for a particle diffusing according to Eq.~(\ref{eq:prop_kin}) with the mass set as the physical mass, $M=M_P$. The (red) circles are the GFMC results for $\omega_c^s \simeq 449$ MeV. }
\label{fig:kinetic}
\end{figure}

In Fig.~\ref{fig:kinetic} we plot the mean square displacement as a function of the
imaginary time for a cutoff of $\omega^s_c \simeq 449$ MeV, and also the curve we would expect from a diffusion given by Eq.~(\ref{eq:prop_kin}) with the physical mass $M_P$. A linear fit to the functional form we propose yields masses that differ by $\sim$ 2 MeV at most from the physical mass, for every cutoff we considered. Thus, in our simulations we set the kinetic mass counter term
to zero $\beta_K=0$, in agreement with our nonrelativistic calculation reported in Appendix \ref{app:selfen}, which shows this correction is small.

\begin{figure}[!htb]
\centering
\includegraphics[angle=-90,width=\linewidth]{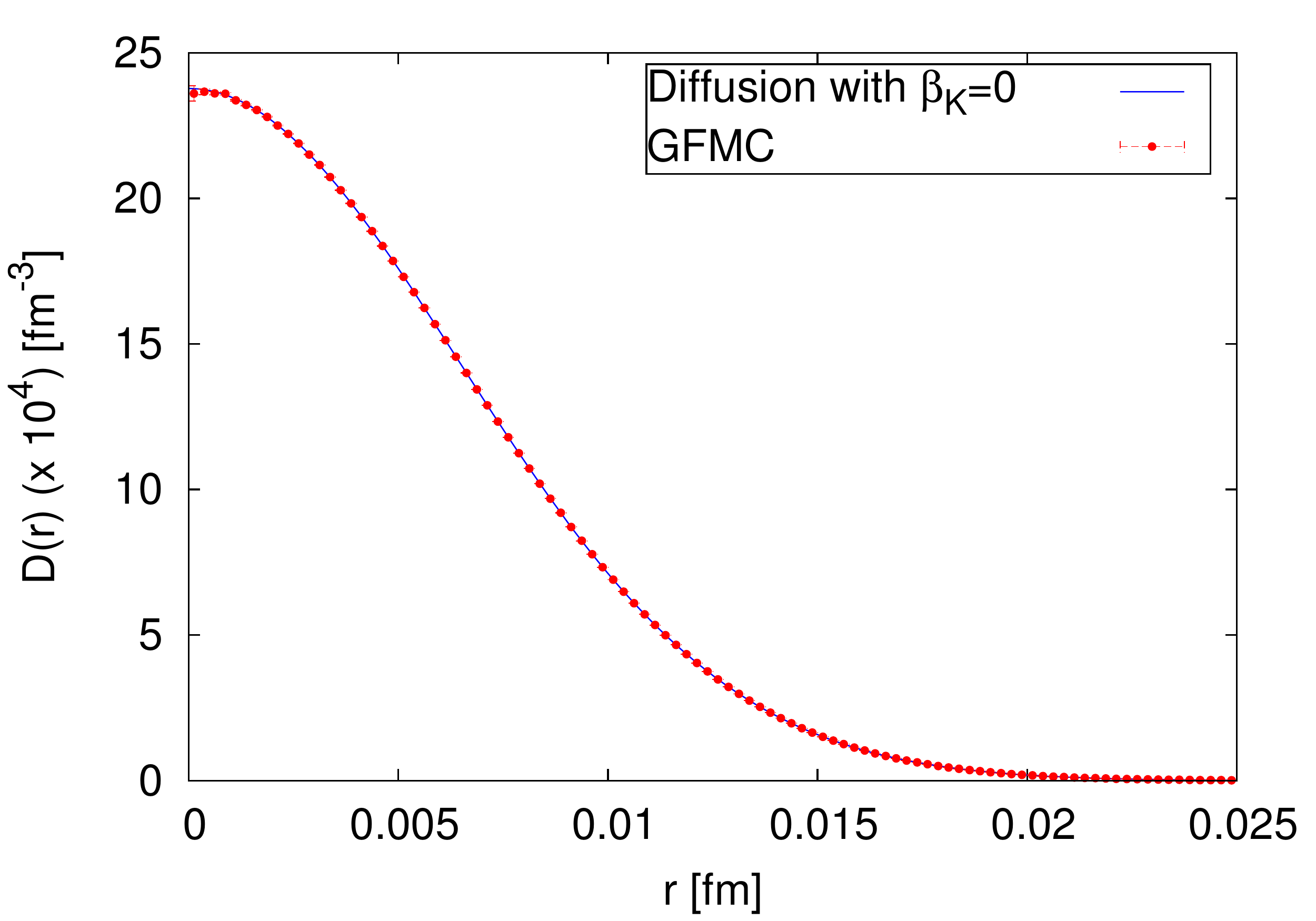}
\caption{(Color online) Euclidean time density correlation as a function of the displacement for a cutoff $\omega_c^s\simeq 449$ MeV. The red circles correspond to the GFMC results, while the blue curve stands for Eq.~(\ref{eq:prop_kin}) evaluated at $M=M_P$.}
\label{fig:dcorr}
\end{figure}

Another way of verifying that we can set $\beta_K=0$
is to calculate
the Euclidean time density correlation function \cite{fet03}, defined as
\begin{eqnarray}
D(\bvec{r})=\frac{\bra{\Psi_T}\rho(\bvec{r})e^{-(H-E_T)\delta \tau}\rho(0)\ket{\Psi_0}}{\product{\Psi_T}{\Psi_0}},
\end{eqnarray}
which
accounts for the nucleon displacement in between diffusion steps. 
In Fig.~\ref{fig:dcorr} we compare our results with the free-particle propagator of Eq.~(\ref{eq:prop_kin}), where we set $M=M_P$, with those obtained
from $D(\bvec{r})$, assuming that the latter is a function of only $r=|\bvec{r}|$, which is true for large enough systems. The fact that in the short-time
limit the nucleon is diffusing with a constant related to $M_P$ is consistent with our GFMC and perturbation theory results.

\begin{figure}[!htb]
\centering
\includegraphics[angle=-90,width=\linewidth]{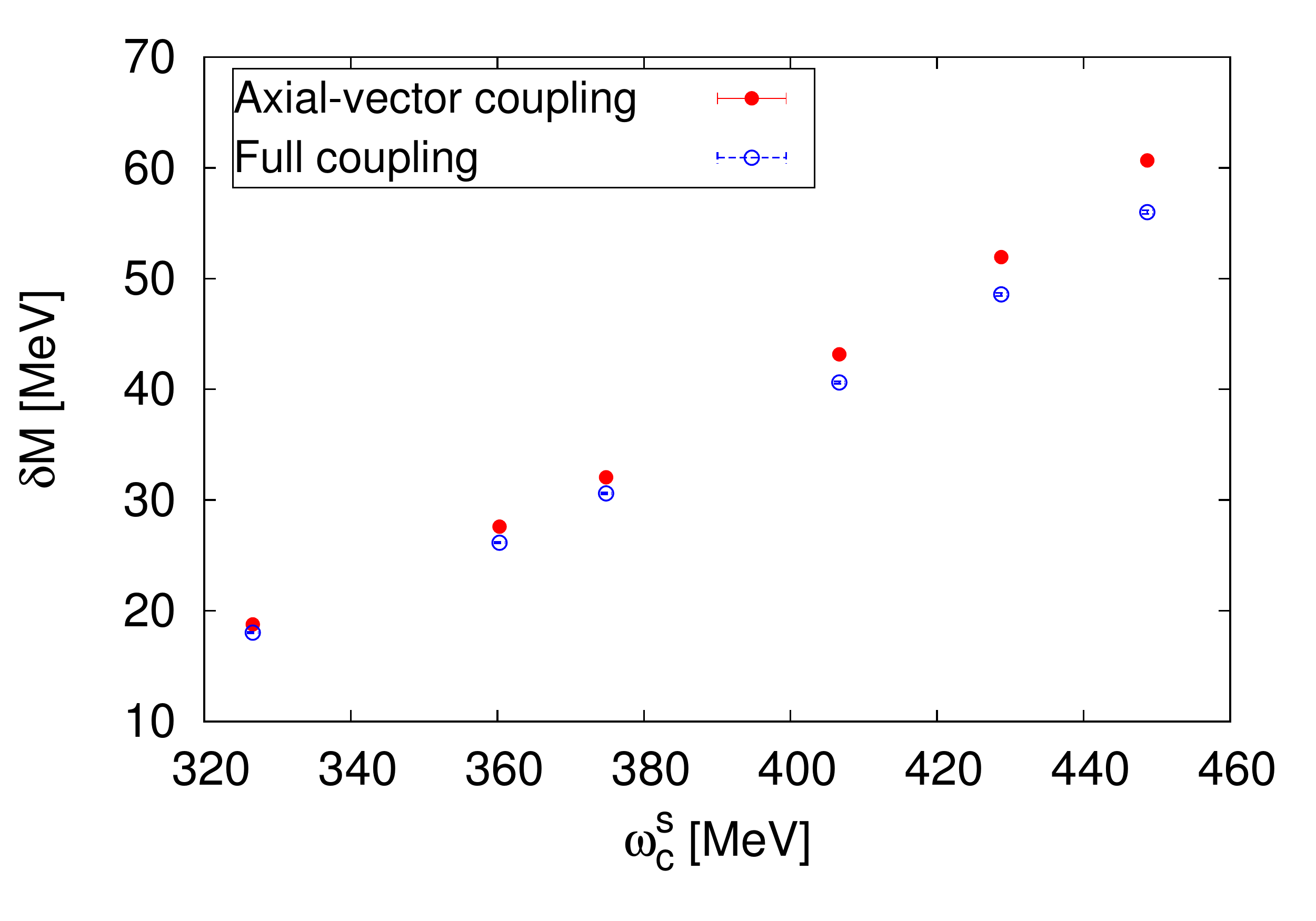}
\caption{(Color online)
Rest mass counter term as a function of the cutoff $\omega_c^s$.
The (blue) open circles are the results with the
full one-nucleon Hamiltonian Eq.~(\ref{eq:hqmc}). The (red) closed circles
are the results neglecting the Weinberg-Tomozawa terms $H_{WT}$.
}
\label{fig:rest}
\end{figure}

The rest mass counter term $\delta M$ is calculated by requiring that the total energy of a single nucleon interacting with the pion field is equal to the physical mass of the nucleon. We investigated
the full single-nucleon Hamiltonian and the one without the Weinberg-Tomozawa term, using the corresponding propagators. We summarize our results in
Fig.~\ref{fig:rest} that are obtained for $L=$ 10 fm. The difference between the mass counter terms is $\simeq 4.7$ MeV for the largest cutoff considered, order 0.5\% of the total rest mass.
Given the simplification in the computational procedures, such small energy difference suggests that it is quite safe to propagate the configurations using the axial-vector coupling only
and computing the Weinberg-Tomozawa contribution to the energy as a first-order perturbation.

\begin{center}
\begin{figure}[!htb]
\centering
\includegraphics[angle=-90,width=\linewidth]{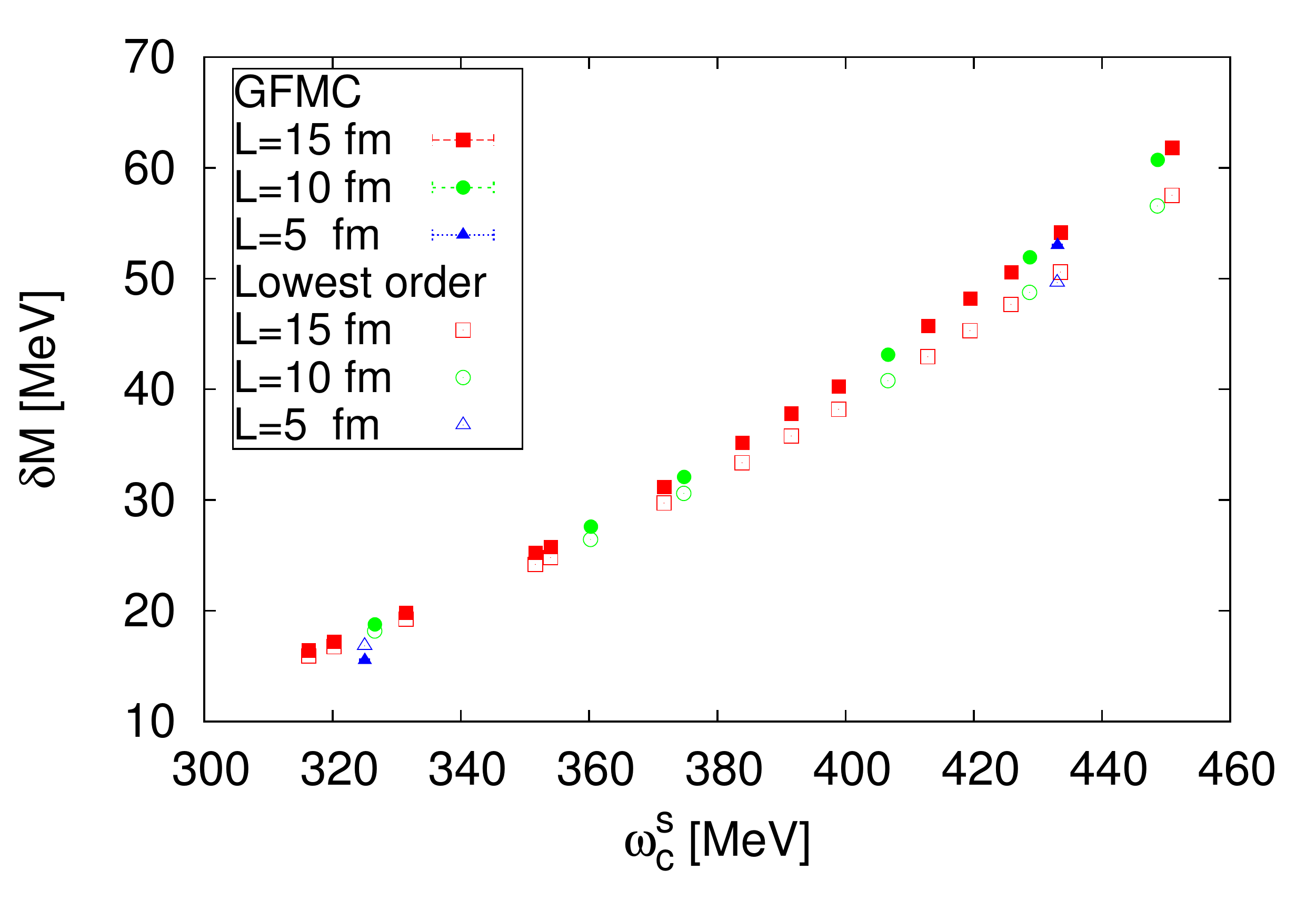}
\caption{(Color online)
The rest mass counter term as a function of the cutoff for $L=5$,
10, 15 fm, (blue) triangles, (green) circles, (red) squares,
respectively. The closed symbols represent GFMC results obtained
discarding $H_{WT}$ in Eq.~(\ref{eq:hqmc}) in
the one-nucleon Hamiltonian. The open
symbols stand for the lowest-order nonrelativistic rest mass calculated
with Eq.~(\ref{eq:selflowp}).}
\label{fig:box}
\end{figure}
\end{center}

We also investigated the dependence of our results on the simulation box size. We varied the side of the box $L=5$, 10, 15 fm,
and we compared the results for the rest mass counter term, neglecting the Weinberg-Tomozawa term $H^{1N}_\text{WT}$. In Fig.~\ref{fig:box} it is possible to see that the counter term calculated with $L=5$ fm deviates from the other values for the smallest cutoff considered. However, the difference between the results obtained with $L=10$ and 15 fm is $\simeq 0.5$\% of $M_P$ at most. Therefore, to speed up the calculations, we chose $L=10$ fm for all the calculations presented in the remainder of this work. As an example, for
$\omega_c^s \sim 449$ MeV, the  box with $L=15$ fm requires more than 3 times the number of $\bvec{k}$ vectors. In Fig.~\ref{fig:box} we also show our lowest order nonrelativistic results for the rest mass counter term, described in Appendix \ref{app:selfen}. The results differ, at most, by 0.4\% of $M_P$.

\subsection{The pion cloud}

One of the most interesting properties that can be computed within the 
formalism presented in this paper are those of the virtual pions 
surrounding the nucleons. 
Although this might in principle contain some dynamical information, at 
present we limit ourselves to analyze static properties.
This calculation is not intended to be a rigorous study of the pion
cloud or nucleon form factors. Instead, we evaluate properties
that demonstrate that our formalism can compute
correlations with explicit pion dependence.

An interesting quantity to analyze is the ground-state momentum distribution of the pion cloud for the different charged states $n_\alpha(\bvec{k})$. Since the sums of Eq.~(\ref{eq:fourier})
are written in such a way that $\bvec{k}$ is included and $-\bvec{k}$ is not, this is best represented by the expectation value of 
\begin{equation}
N_{i\bvec{k}} = a^\dagger_{\alpha\bvec{k}} a_{\alpha\bvec{k}} +  a^\dagger_{\alpha\bvec{-k}} a_{\alpha\bvec{-k}},
\end{equation}
with the creation and annihilation operators for a pion in a given charge state are given in Eqs.~(\ref{eq:charged_a}) and (\ref{eq:neutral_a}).  We computed the momentum distributions and radial densities of the pion cloud using the forward walking procedure described in Sec.~\ref{sec:qmc}
to avoid the bias due to the trial wave function. We considered a box with $L$=10 fm, and the model state $\ket{\Phi}$ of Eq.~(\ref{eq:psitqmc}) corresponding to a spin-up proton.

\begin{figure}[!htb]
\centering
\includegraphics[angle=-90,width=\linewidth]{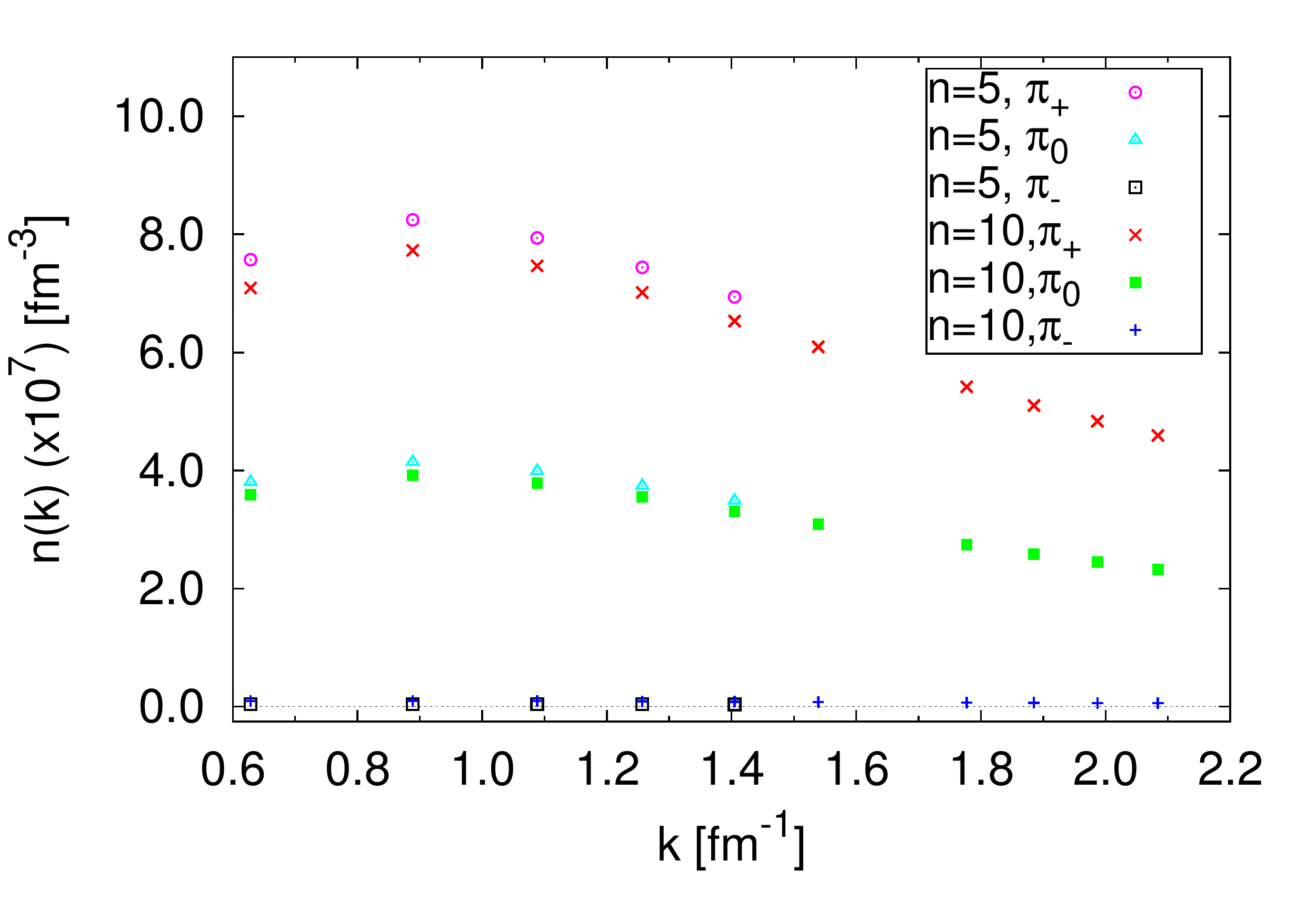}
\caption{(Color online)
Momentum distribution for the different charge states, for
systems with different shell numbers, $n=5$ corresponds to a cutoff $\omega_c^s\simeq 327$ MeV, and 
$n=10$ to $\omega_c^s\simeq 449$ MeV. The different symbols correspond to $\pi_+$, $\pi_0$, and $\pi_-$ for $n=5$, 
(purple) open circles, (cyan) triangles, and (black) open squares, respectively; and $\pi_+$, $\pi_0$, and $\pi_-$ for $n=10$,
(red) crosses, (green) solid squares, and (blue) pluses, respectively.
The $y$-axis extends a little lower than zero (dotted line)
only to allow a clear picture for the $\pi_-$ distributions,
the distributions are never negative.
}
\label{fig:mdist}
\end{figure}

In the limit $L\to\infty$, $n_\alpha(\bvec{k})$ should be a function of $k=|\bvec{k}|$ alone. Already for $L$=10 fm we found minimal differences among the modes with the same $k$, hence in Fig.~\ref{fig:mdist} we show the pion momentum distribution as a function of $k$, only.
The normalization is chosen such that $N_\alpha=L^3\sum_i n_\alpha(k_i) g_i$, where $N_\alpha$ is the total number of pions of charge $\alpha$, and $g_i$ is the multiplicity of the $i$-th shell. An interesting feature is that the distribution of $\pi_+$ is approximately twice the one of $\pi_0$. This follows from the structure of the axial-vector coupling, which involves
\begin{equation}
\label{eq:taudotpi}
\tau_i\pi_i=\frac{1}{2}\tau_+(\pi_x - i\pi_y)+\frac{1}{2}\tau_-(\pi_x + i\pi_y)+\tau_z \pi_0,
\end{equation}
with $\tau_\pm=(\tau_x\pm i\tau_y)$ being the isospin raising and lowering operators, and $\pi_0=\pi_z$. If we suppose that the cartesian $\pi_i$ are produced in the same amount, then we expect twice as many $\pi_0$ than $\pi_+$. Since we are looking at a one proton state, the production of $\pi_-$ is much smaller compared to that of $\pi^+$ and $\pi^0$. Conversely, if the baryon is a neutron, we get analogous results with the distributions of $\pi_+$ and $\pi_-$ interchanged. Although increasing the cutoff increases the total pion production, the number of pions at low-momenta appears to be cutoff independent. 

\begin{figure}[!htb]
\centering
\includegraphics[angle=-90,width=\linewidth]{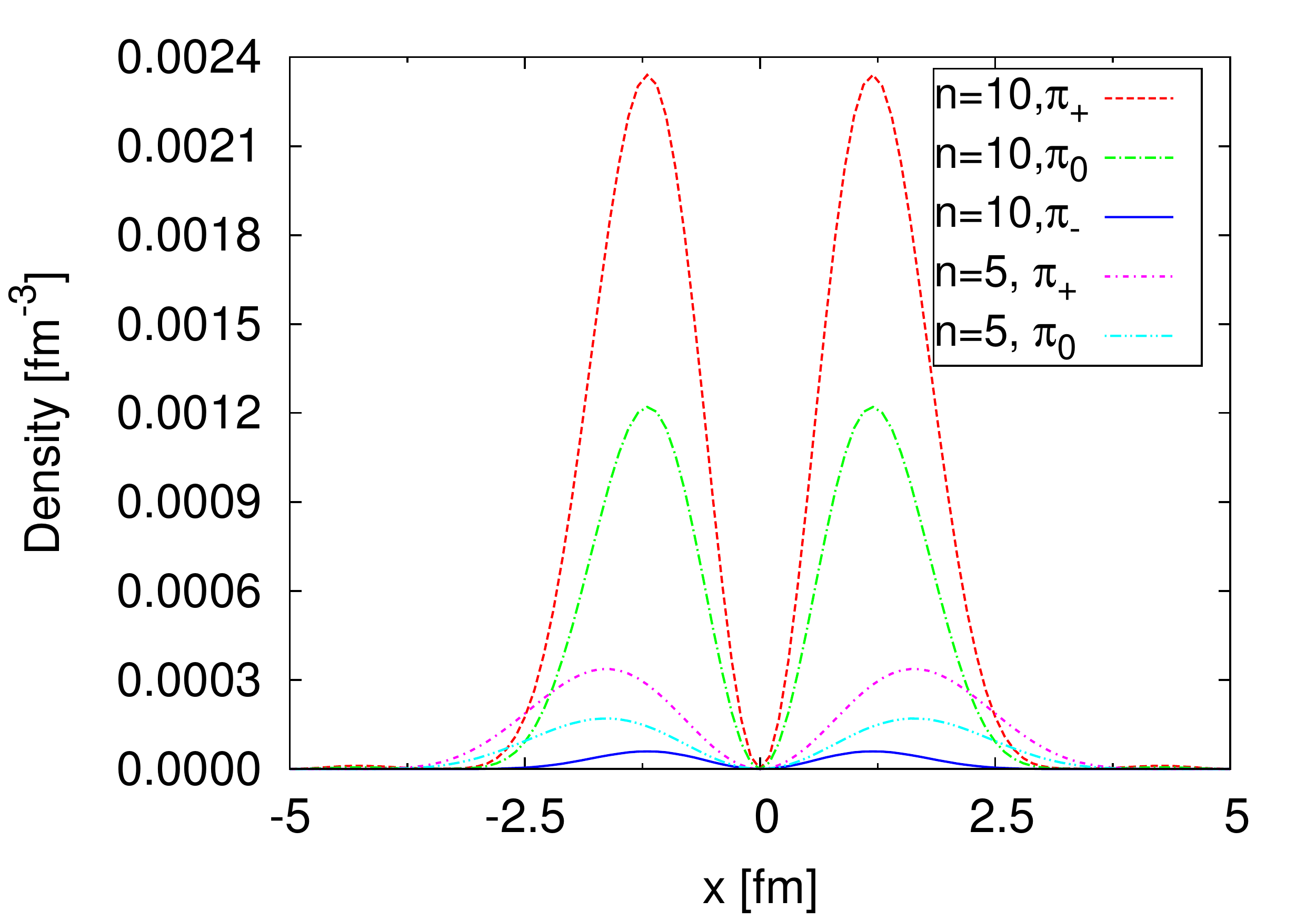}
\caption{(Color online)
Pion density for the different charge states as a function of $x$
coordinate of the box.
We plot the density for two systems with different shell numbers,
$n=5$ corresponds to a cutoff $\omega_c^s\simeq 327$ MeV, and 
$n=10$ to $\omega_c^s\simeq 449$ MeV. The different curves correspond to $\pi_+$, $\pi_0$, and $\pi_-$ for $n=10$, 
red long-dash, green dot dash, and solid (blue), respectively; and $\pi_+$, $\pi_0$ for $n=5$, purple dot short-dash, and
cyan double-dot dash, respectively.
}
\label{fig:pdensity}
\end{figure}

The pion densities, whose off-diagonal components are related to the momentum distributions through a Fourier transform, can also be resolved for different charge states, as in Eq.~(\ref{eq:dens_charge}). The results for the density are displayed in Fig.~(\ref{fig:pdensity}) for a spin-up proton as model state -- we did not plot the $n=5$ density for $\pi_-$ because it is negligible in the scale of the figure. In analogy to $n_\alpha(k)$, the production of $\pi_-$ is heavily suppressed. If the model state is a neutron, we, of course, get identical results with the densities of $\pi_+$ and $\pi_-$ interchanged.

\subsection{One pion exchange}
As mentioned above, the long-range behavior of the nuclear force is
due to the one-pion exchange. It arises from tree-level diagrams
with four external nucleons and an off-shell pion. At lowest order in
perturbation theory, the potential arising from two static nucleons is
\begin{eqnarray}
\label{eq:OPEpot}
V_{\rm OPE} (\bvec q)=
-\left( \frac{g_A}{2f_\pi}\right)^2\frac{(\bvec \sigma_1\cdot \bvec q)
(\bvec \sigma_2\cdot\bvec q)}{q^2+m_\pi^2} \bvec \tau_1\cdot \bvec \tau_2,
\end{eqnarray}
where $\bvec{q}$ is the transferred momentum.
The coordinate-space potential is recovered from $V_{\rm OPE}(\bvec{q})$
via a Fourier transform. To make a meaningful comparison we
need to compute the one-pion exchange potential keeping into account
the geometry and
the cutoff of the simulation cell we use.

In Eq.~(\ref{eq:hfixed}) the last term on the right-hand-side of the fixed nucleon
Hamiltonian contains contributions of the self-energy of the nucleons
and the one-pion exchange potential, in which we are interested. Keeping
only terms that involve the coupling between the two nucleons,
\begin{align}
\label{eq:OPE}
V_{\rm OPE}(\bvec r)&=-\frac{1}{L^3}
\frac{g_A^2}{2f_\pi^2}
\bvec \tau_1 \cdot \bvec \tau_2
{\sum_{\bvec k}}^\prime
(\bvec \sigma_1\cdot \bvec k) (\bvec \sigma_2\cdot \bvec k)\nonumber\\
&\times \frac{\cos(\bvec k\cdot \bvec r)}{\omega_k^2},
\end{align}
which is consistent with Eq.~(\ref{eq:OPEpot}).

\begin{figure}[!htb]
  \includegraphics[angle=-90,width=\linewidth]{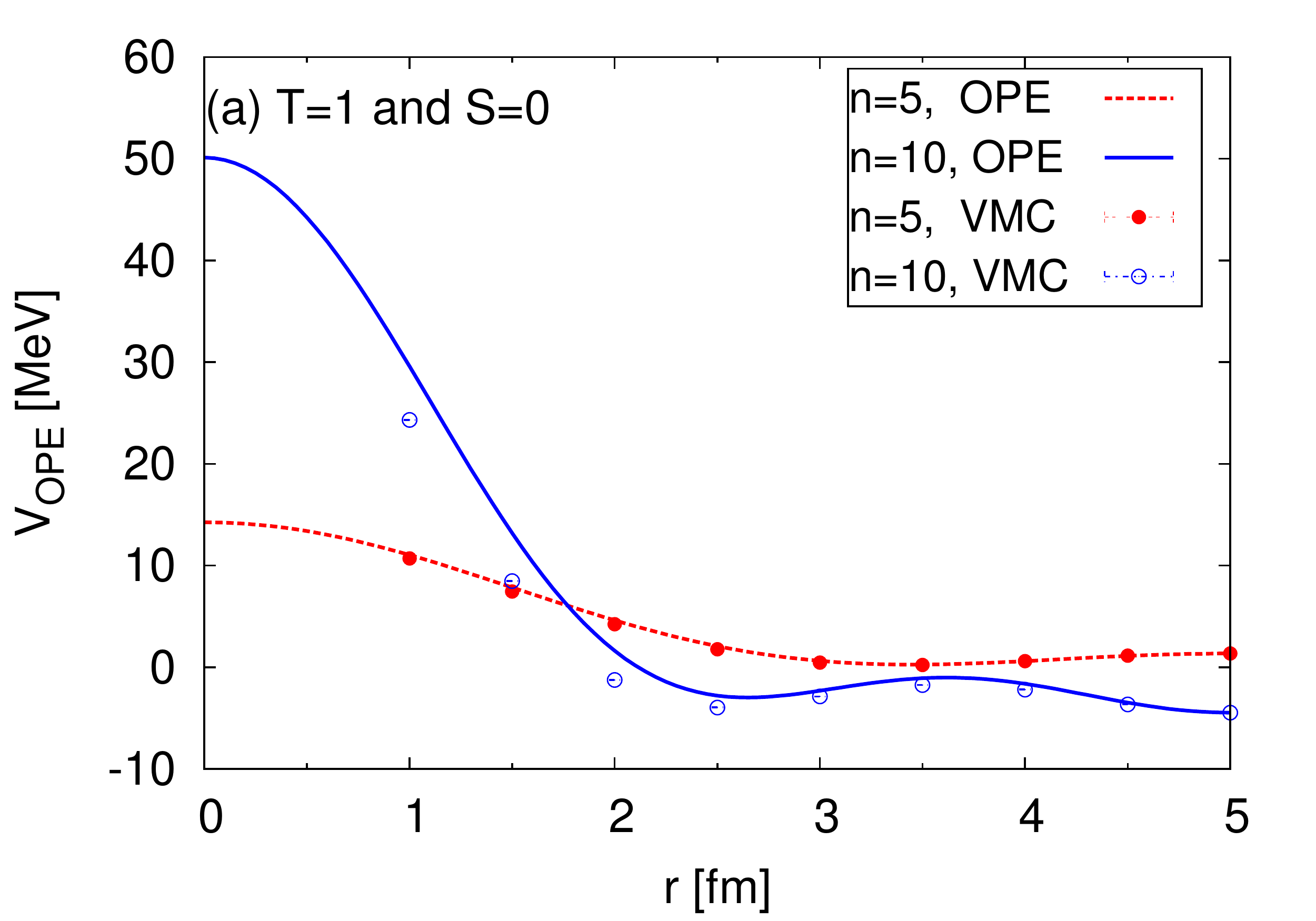}
  \includegraphics[angle=-90,width=\linewidth]{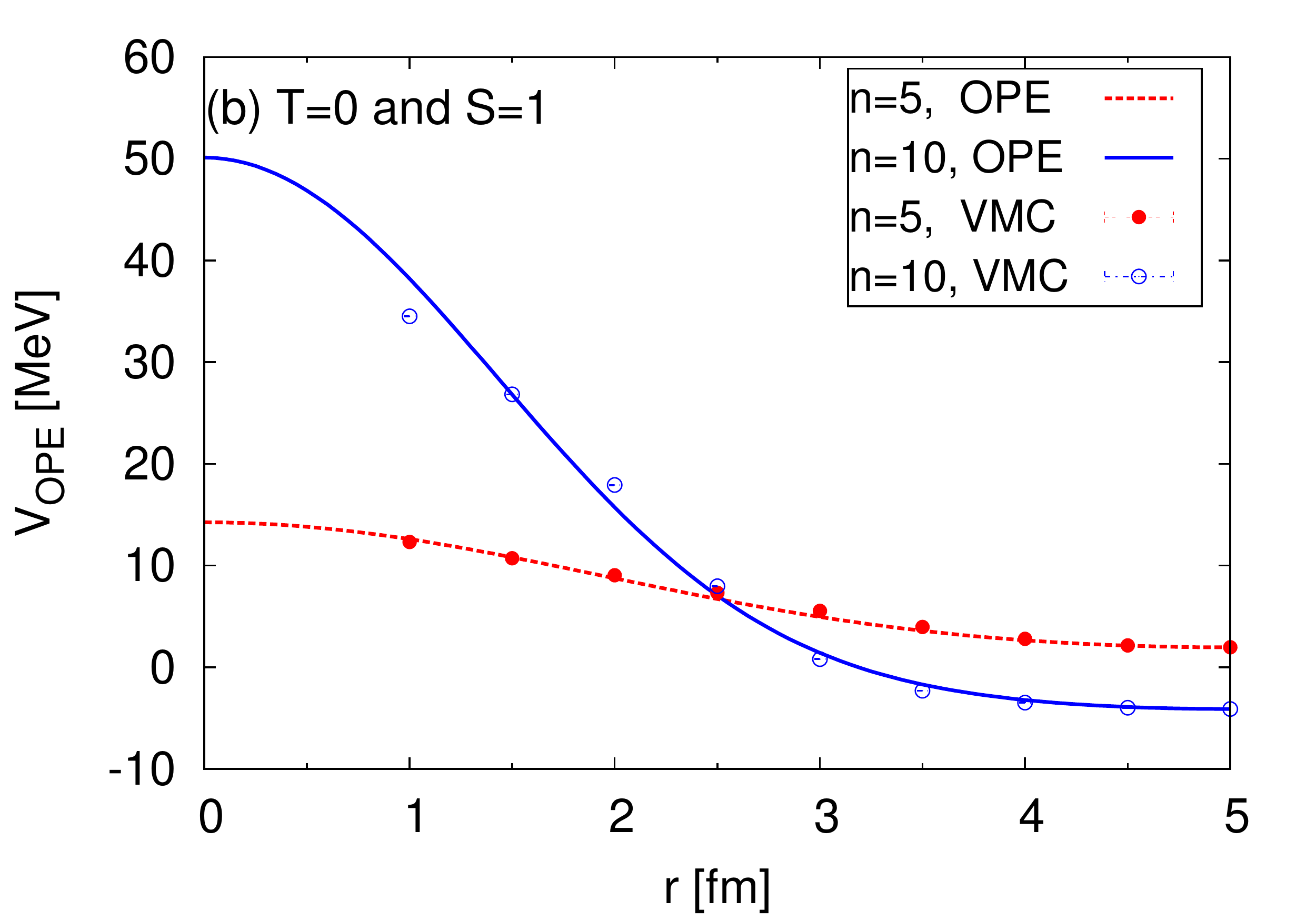}
\caption{(Color online)
One-pion exchange potential for two nucleons a distance $r$ apart along
the x-axis in the $T=1$ and $S=0$ channel (upper panel) and $T=0$ and
$S=1$ channel (lower panel). The points (VMC) correspond to our variational
results, where the full (red) circles denote $n=5$ ($\omega_c^s\simeq 327$
MeV) and open (blue) circles stand for $n=10$ ($\omega_c^s\simeq 449$
MeV). The curves (OPE)
correspond to the one-pion exchange potential of Eq.~(\ref{eq:OPE})
with the same cutoff as the VMC calculations.}
\label{fig:ope}
\end{figure}

The instantaneous one-pion exchange potential neglects terms where
two or more pions are exchanged and the vertices are in different time orders.
These commutator terms contribute even for fixed nucleons.
However, they become unimportant for large nucleon separations.
We studied the interaction between two fixed nucleons as a function
of the inter-particle distance $r$ in the $T=1$ and $S=0$ and $T=0$
and $S=1$ channels. We used VMC calculations and checked that they
were accurate by performing GFMC calculations at a few separations.
Our VMC results, represented by the points
in Fig.~\ref{fig:ope}, are obtained by subtracting the nucleon
self-interaction terms from the ground-state expectation value of
$H_{\pi\pi}+H_{AV}$
for two different spherical cutoffs. For comparison, we also
show the curves corresponding to the one-pion exchange
potential of Eq.~(\ref{eq:OPE})
for the same cutoff employed in the VMC calculations. As expected,
the VMC results agree with the one-pion exchange potential at sufficiently large
distances, $r\gtrsim 3.0$ fm.
The differences at smaller distances
come from the fact that we are solving for a Hamiltonian that contains
terms other than the one-pion exchange.
This is one of the key
features of explicitly including the modes of the pion field, which is
absent in potential models, in which
multiple pion-exchange potentials have to be explicitly devised.

\subsection{Two nucleons}
\label{sec:2nuc}
We need to fix the low-energy constants $C_S$ and $C_T$ associated to the contact terms entering $H_{NN}^{2N}$ of Eq.~(\ref{eq:h_nn2n}). These should be
either fitted to experiment or to QCD. Instead of fitting to experiment, we
take the expedient step of fitting to results of a potential model that
has been fit to experiments. Since our calculations rely on a periodic box,
we fit $C_S$ and $C_T$ to reproduce the ground state
results of the Argonne $v_6^\prime$ (AV6P)
potential\cite{wir95} for the deuteron and two-neutrons in a periodic box.

Note that
a possible way to directly fit experiments would involve the
L\"uscher method \cite{lus91}.
The energy spectrum of a system of two particles in a box with 
periodic boundary conditions, for box sizes greater than the 
interaction range, and for energies below the inelastic threshold, 
is
determined by the scattering phases at these energies.
The L\"uscher method can be used to compute the energy levels 
given the scattering phases or, conversely, to calculate the 
scattering phases if the energy spectrum is known.
For larger systems ($A\geqslant 3$)
the concepts and ideas
concerning finite-volume corrections
to the binding energy of an
$A$-particle bound-state in a periodic box, developed
in Ref.~\cite{kon18},
might be useful.

\begin{figure}[!htb]
  \includegraphics[angle=-90,width=\linewidth]{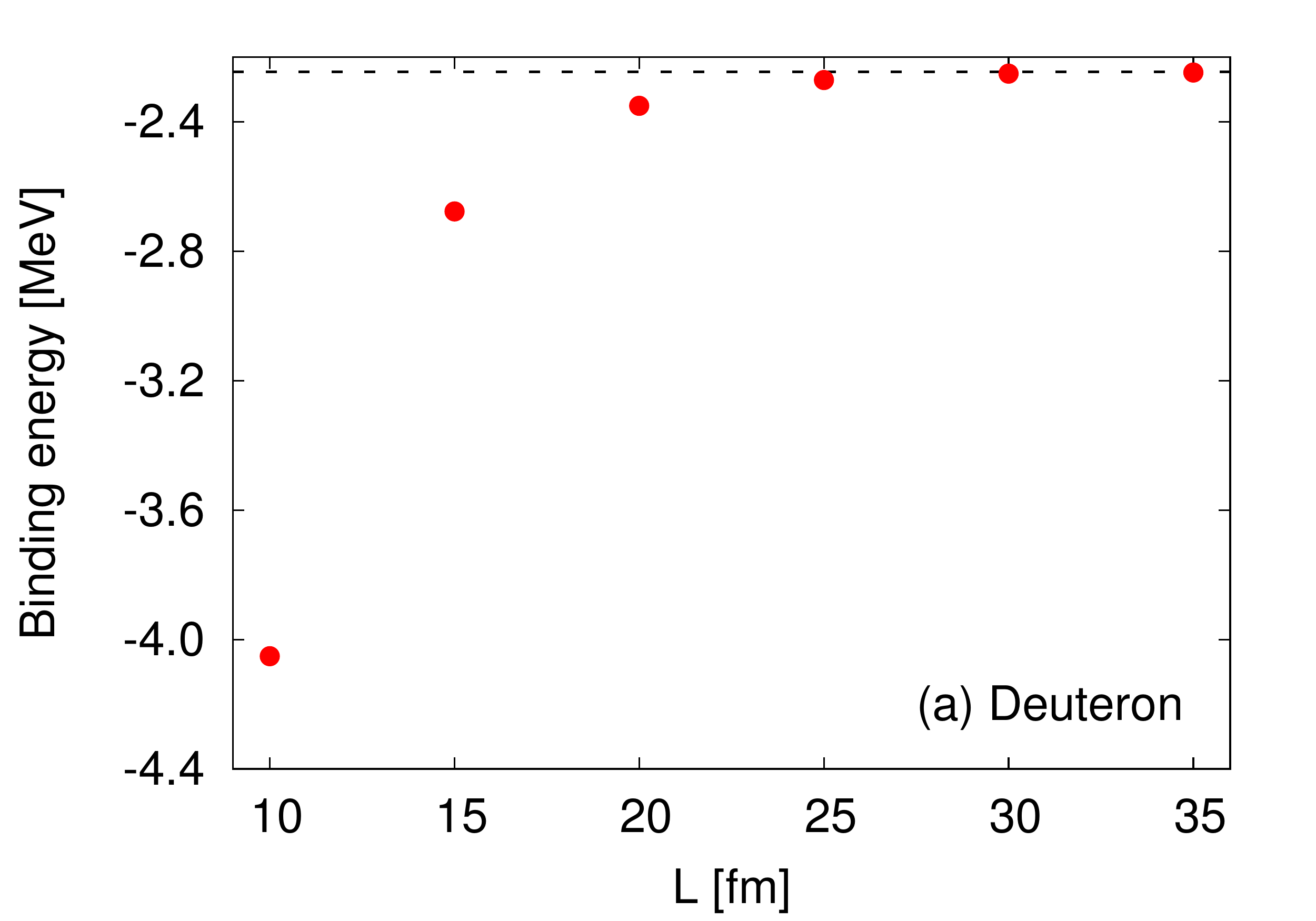}
  \includegraphics[angle=-90,width=\linewidth]{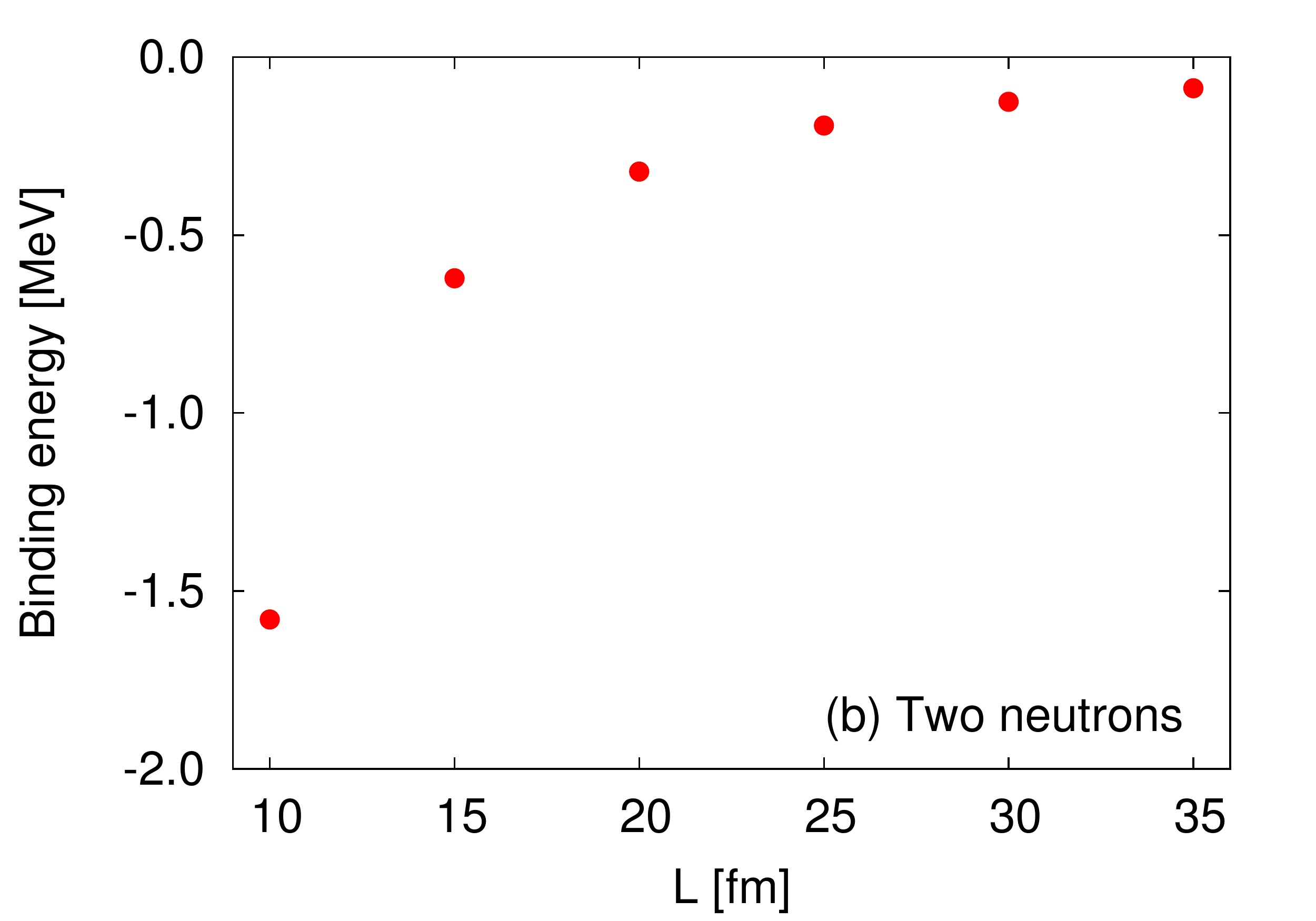}
  \caption{Binding energy of the deuteron (upper panel) and two neutrons (lower panel) in a box. The symbols correspond to the energy calculated using the
Lanczos algorithm with the AV6' potential.
The dashed line corresponds to the binding energy of the deuteron in free space using the AV6'
potential.\label{fig:2nucleons}}  
\end{figure}

As a first step, we developed a numerically stable version of the Lanczos algorithm \cite{lan50} to solve for the energy
of the deuteron and two neutrons in a periodic box using the AV6P potential
and a plane wave basis.
By imposing
periodic boundary conditions, the continuum version of the AV6P potential, which has the operator structure of Eq.~(\ref{eq:v(r)}), is modified to include periodic images from the surrounding boxes,
\begin{equation}
V_{NN}(\bvec{r}_{12}) \to \sum_{\bvec{n}} V(\bvec{r}_{12}+L\bvec{n}),
\label{eq:periodic_v}
\end{equation}
where $\bvec{n}=(n_x,n_y,n_z)$ with $n_i$ integers numbers. The self
potential energy term of the periodic images is included.
We showed that for $L\geqslant$ 10 fm one image
in each direction is sufficient to obtain periodic solutions since the AV6P interaction is at most of pion range.
In Figs.~\ref{fig:2nucleons}(a) and (b) we plot the binding energy of the deuteron and two neutrons,
respectively as a function of the box side. For $L\leqslant25$ fm, the deuteron energies are much lower than the 
value for the system in free space. However, for $L\geqslant 25$ fm the agreement between finite periodic box results 
and the continuum is remarkably good. 

We then tune $C_S$ and $C_T$ in the GFMC simulations with explicit 
pions to reproduce the
energies of both two nucleon systems. We do not include the 
Weinberg-Tomozawa term, as the one-nucleon 
results suggest it will provide a small contribution for the 
momentum cutoffs we employed. Based on the results
of Fig.~\ref{fig:2nucleons}, we performed the explicit-pion 
calculations only for $25 \leqslant L \leqslant 35$ fm. 
The pion nucleon axial-vector coupling in our formalism is already 
periodic, and so are the contact terms using Eq.~(\ref{eq:deltakc}),
hence we do not need to modify them with 
Eq.~(\ref{eq:periodic_v}).

\begin{figure}[!htb]
   \includegraphics[angle=-90,width=\linewidth]{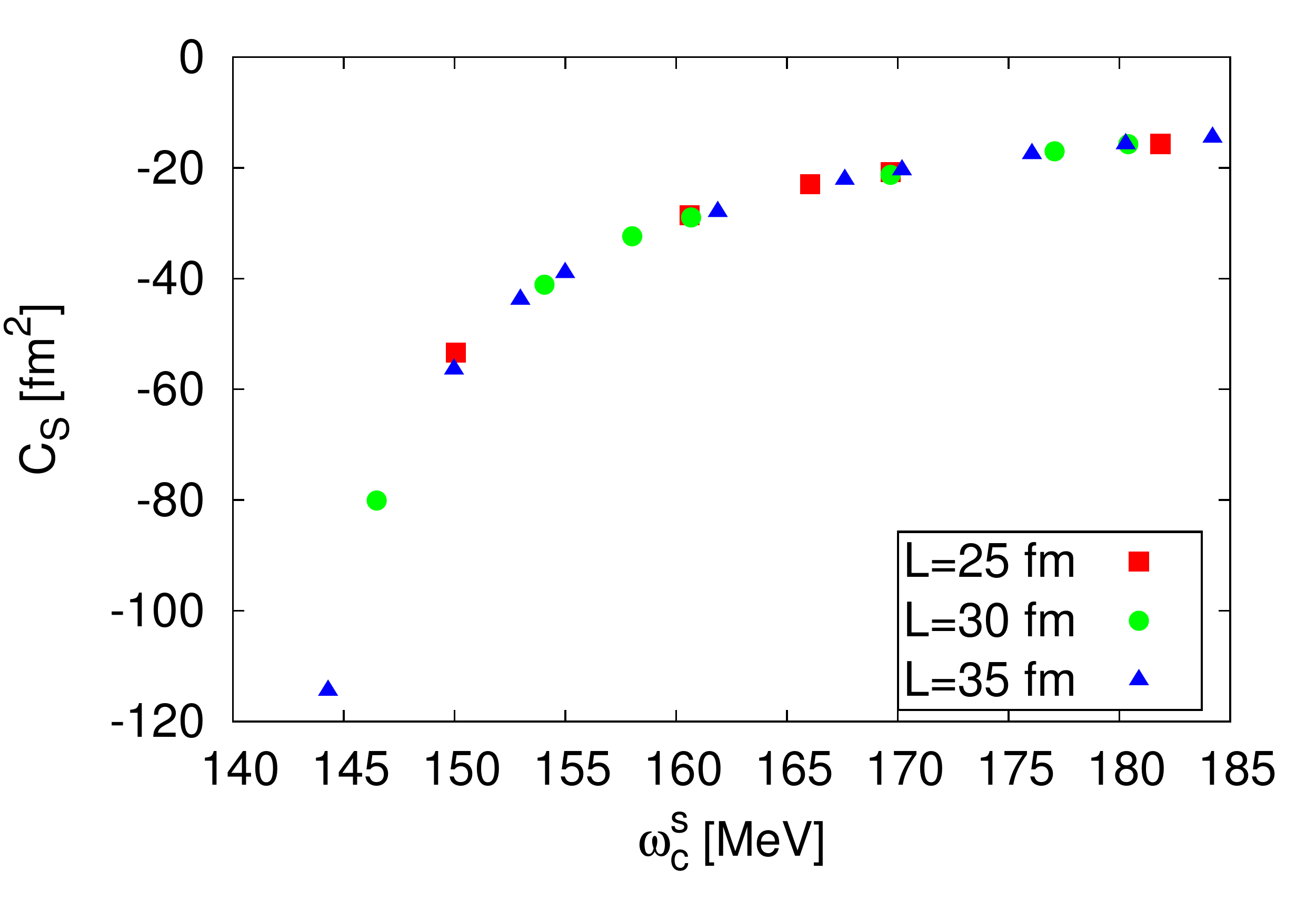}
   \includegraphics[angle=-90,width=\linewidth]{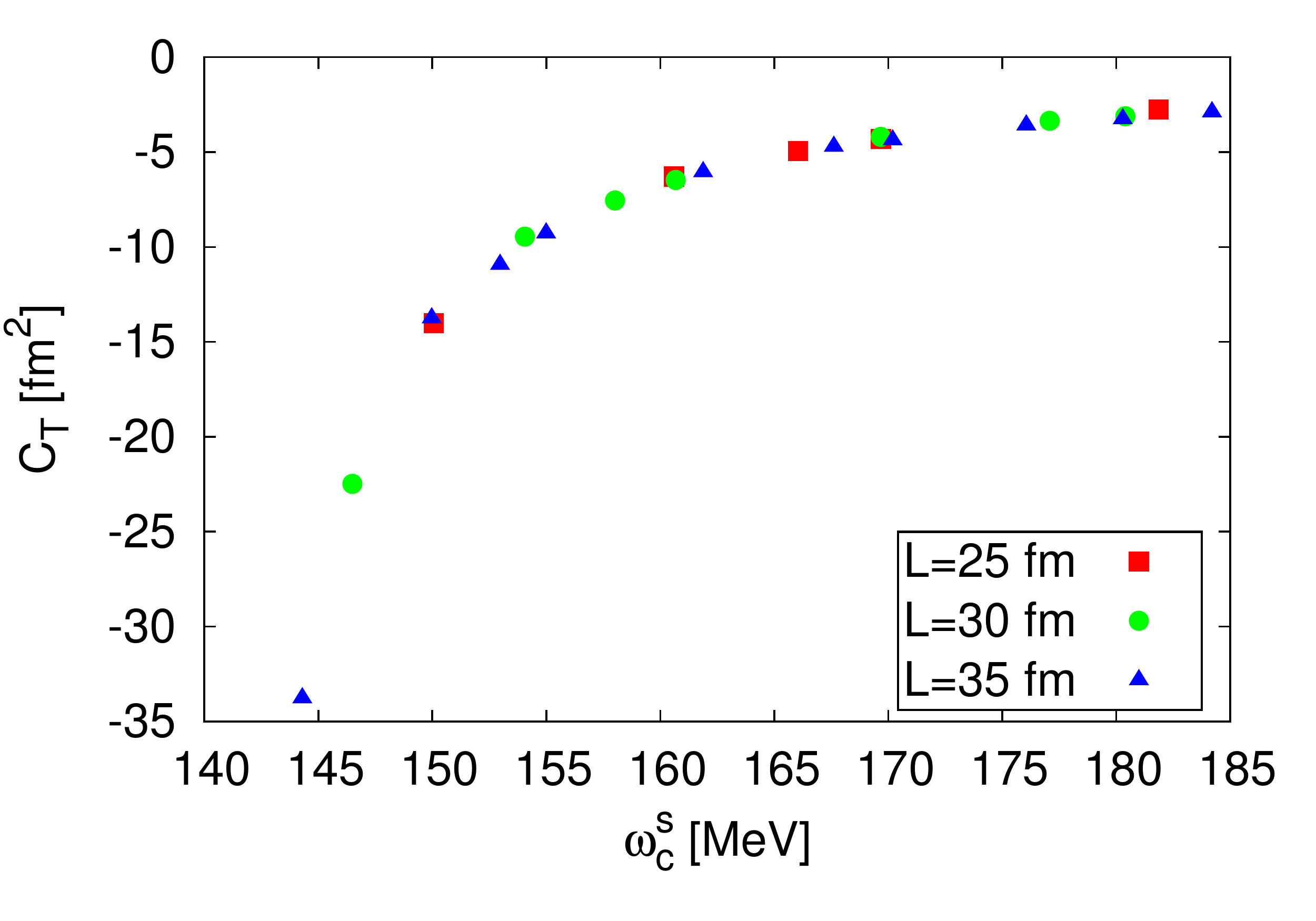}
   \caption{(Color online)
Low-energy constants $C_S$ (upper panel) and $C_T$ (lower panel)
as a function of the cutoff $\omega_c^s$
for $L=25$, 30, and 35 fm,
(red) squares, (green) circles, and (blue) triangles, respectively.
The values are reported in Table~\ref{tab:contact}.\label{fig:csct}}  
\end{figure}

The fitted values of $C_S$ and $C_T$  for different box sizes and cutoffs are 
shown in Fig.~\ref{fig:csct} and
reported 
in Table~\ref{tab:contact}.
We are aware that the cutoffs we used are very low compared to
those typically used in other chiral EFT formulations. This choice is by no means
due to an intrinsic limitation of the method, but to the extent of the
computational effort that we deemed reasonable to obtain these demonstrative results.
In Appendix~\ref{app:contact} we report values of these LECs for
a different choice of the contact interaction.

\begin{table*}[!htb]
\centering
\caption{
Contact parameters for different box sizes, $L$=25, 30, and 35 fm, as a function of the cutoff $\omega_c^s$.
The $\omega_c^s$ are given in MeV, while $C_S$ and $C_T$ are in fm$^2$.}
\begin{tabular}{|c|c|c|c||c|c|c||c|c|c|}
\hline
\multicolumn{1}{|l|}{} &
\multicolumn{3}{c||}{$L$= 25 fm} &
\multicolumn{3}{c||}{$L$= 30 fm} &
\multicolumn{3}{c|}{$L$= 35 fm} \\ \hline
\multicolumn{1}{|l|}{n} &
$\omega_c^s$ & $C_S$ & $C_T$ &
$\omega_c^s$ & $C_S$ & $C_T$ &
$\omega_c^s$ &$C_S$ & $C_T$ \\ \hline\hline
1 & 150.06 & -53.35 & -14.02 & 146.49 & -80.09 & -22.48 & 144.30 & -114.46 & -33.78\\ \hline
2 & 160.61 & -28.56 & -6.29  & 154.06 & -41.11 & -9.46  & 149.98 & -56.41 & -13.76\\ \hline
3 & 166.05 & -22.99 & -4.95  & 158.02 & -32.34 & -7.55  & 152.97 & -43.82 & -10.93\\ \hline
4 & 169.68 & -20.73 & -4.32  & 160.67 & -28.95 & -6.47  & 154.99 & -38.99 & -9.28\\ \hline
5 & 181.85 & -15.68 & -2.75  & 169.67 & -21.25 & -4.20  & 161.88 & -27.96 & -6.06\\ \hline
6 &        &        &        & 177.08 & -16.99 & -3.35  & 167.61 & -22.14 & -4.71\\ \hline
7 &        &        &        & 180.40 & -15.72 & -3.11  & 170.19 & -20.44 & -4.38\\ \hline
8 &        &        &        &        &        &        & 176.06 & -17.51 & -3.59\\ \hline
9 &        &        &        &        &        &        & 180.28 & -15.70 & -3.26\\ \hline
10&        &        &        &        &        &        & 184.20 & -14.54 & -2.90\\ \hline
\end{tabular}
\label{tab:contact}
\end{table*}

\section{Summary and outlook}
\label{sec:summary}

In this paper we describe a promising scheme to explicitly include pion fields
in a Quantum Monte Carlo calculation of a one- and two-nucleon systems. 
This approach can be readily extended to larger nuclei, consistently with the
limits of application of the underlying GFMC (or AFDMC) techniques. 
One important remark to be made is that, since pion fields are bosonic,
no further contribution to the fermion sign/phase problem is introduced.

The first application to the one-nucleon system is meant to 
verify the consistency of the method itself. In particular, we analyzed finite-size effects,
and the extent of the differences due to the choice of the initial Lagrangian.
We first studied the renormalization of the nucleon mass with a
Hamiltonian in which the coupling between the nucleons and the pion fields
is described by an axial-vector interaction. A consistency check against
first-order diagrammatic calculation of the self-energy of the nucleon has been
successfully carried out.  We tried to assess the importance of including
the Weinberg-Tomozawa coupling in the interaction. Although this term
appears at leading order in the chiral expansion, we showed that its effect in
the renormalization of the nucleon mass is much
smaller than that of the axial-vector coupling. 
One interesting possibility opened by our method is the direct study
of the pion distribution.
In the one-nucleon sector, we analyzed the momentum and density distributions of the pion cloud surrounding the nucleon. 
Although many details are still missing, this can be thought of as a first
step towards the calculation of the single-nucleon electroweak form
factors. Standard chiral-EFT calculations fail to describe the proton
and nucleon form factors for momentum transfers beyond $Q^2 \sim 0.1$
GeV$^2$~\cite{Kubis:2000zd}. The inclusion of vector mesons sensibly
improve the agreement with data~\cite{Scherer:2009bt}. Within our
explicit-pion QMC framework, we plan to assess whether the resummation
of important higher-order contributions can mimic their inclusion.

Turning to the two-body problem, the correct asymptotic behavior of the potential between
two static nucleons was verified. As expected, the short- and
intermediate-range part of the potential differs from the OPE expression,  
due to multiple-pion exchange, automatically included in our formalism.
The low-energy constants of the contact terms in the Hamiltonian were 
determined by fitting exact diagonalization results on the binding energy
of the two-body problem (pn and nn) in a finite box. This is a necessary
step towards the simulations of light nuclei within the explicit-pion
formalism. In this paper we employed a sharp spherical momentum cutoff. 
The dependence of results on the specific choice of the regularization will 
be explored in future works.

The validity
of instantaneous pion interaction approximations in
potential models can be tested by
computing properties of light-nuclei using our formalism.
As previously mentioned, the extension of the calculations to 
larger systems
(and in particular $A$=3 and 4 nuclei) is straightforward,
aside for the larger
computational cost,
and it is currently in progress. 
For $N_\pi$ pion modes and $A$ nucleons, the propagation of
the pions costs $O(N_\pi)$, while the axial-vector and Weinberg-Tomozawa
interactions cost $O(AN_\pi)$. In principle, since these interactions
are local in real space, they can be written using a fast Fourier transform
to be $O(N_\pi\ln N_\pi)+O(A)$, but we expect $A \leqslant \ln N_\pi$ for
most systems. Much of the computational cost of the algorithm presented
here has to be ascribed to the explicit sum over the many-body spin/isospin 
states that we perform in the imaginary-time propagations of the nucleons.
This part of our
algorithm scales exponentially and costs $O(4^A)$. GFMC calculations
scale slightly better since they conserve nuclear charge (and sometimes
isospin) -- for example, instead of $4^A$ spin/isospin states, a nucleus
with good charge
$Z=A/2$ has $\frac{A!}{(A/2)!^2} 2^A \simeq 4^A\sqrt{\frac{2}{\pi A}}$
spin/isospin states.
To go beyond light nuclei, we plan to implement a spin-isospin sampling 
algorithm, analogous to that of AFDMC. The latter includes all charge states 
in its basis, and its computational cost scales polynomially with the number
of nucleons $O(A^3)$ for evaluating simple trial wave functions. For the small
systems used here, GFMC methods give lower variance for the same
computational time.

In analogy to other quantum Monte Carlo other methods, computing 
complicated operator expectations is likely to be computationally more
intensive. For example, the pion distributions as computed here require 
mixed derivatives with respect to the pion coordinates, and these
scale as $O(N_\pi^2)$. This behavior is ameliorated by the fact that these 
calculations are only carried out on uncorrelated samples, not at every step.

\begin{acknowledgments}
We thank Pietro Faccioli, Paolo Armani, Ubirajara van Kolck, and 
Juan Nieves for important 
discussions about the subject of this paper.
This work was supported by the National Science
Foundation under Grant No. PHY-1404405.
This work used the Extreme Science and Engineering Discovery Environment (XSEDE)
SuperMIC and Stampede2
through the allocation TG-PHY160027, which is supported by National Science Foundation Grant No. ACI-1548562.
This research has been supported by the U.S. Department
of Energy, Office of Science, Office of Nuclear
Physics, under Contract No. DE-AC02-06CH11357.
\end{acknowledgments}

\appendix
\section{Conventions}
\label{app:conv}

We use units such that $\hbar=c=1$.
The contravariant space-time and momentum four-vectors are given by 
$x^\mu = (t,\bvec{x})$ and $p^\mu = (E,\bvec{p})$.
Greek indices $\mu,\nu, ...$ run over the four space-time coordinate labels $0,1,2,3$, with $x^0 = t$ being the time
coordinate. Latin indices $i,j,k$, and so on run over the three space coordinate labels $1,2,3$.
The metric is given by
$g^{\mu\nu} = g_{\mu\nu}$ with $g^{00}=1$, $g^{ii}=-1$.  
The covariant versions of the above-mentioned vectors are
$x_\mu = g_{\mu\nu} x^\nu =(t,-\bvec{x})$ and $p_\mu = g_{\mu\nu} p^\nu= (E,-\bvec{p})$.
While for an ordinary three-vector we have $\bvec{x}=(x^1,x^2,x^3)$, 
the three-dimensional gradient operator is defined to be
\begin{equation}
{\nabla} = (\partial_1,\partial_2,\partial_3)
\end{equation}
with
\begin{equation}
\partial_i = \frac{\partial}{\partial x^i} = - \frac{\partial}{\partial x_i} =- \partial^i .
\end{equation}
The Levi-Civita tensor is defined as $\epsilon^{ijk} =1$ if $(i, j, k)$ is an even permutation of $(1, 2, 3)$,
$\epsilon^{ijk} =-1$ if it is an odd permutation and $\epsilon^{ijk} =0$, otherwise.

The spin 1/2 and isospin 1/2 operators of the nucleons are defined as
$\bvec s=\bvec \sigma/2$ and $\bvec t =\bvec \tau/2$, where $\bvec \sigma$
and $\bvec \tau$ are the Pauli matrices operating in spin and isospin space, respectively.
The Pauli matrices are
\begin{eqnarray}
\sigma^1 =
\begin{pmatrix}
0 & 1 \\
1 & 0
\end{pmatrix}
; \ 
\sigma^2 =
\begin{pmatrix}
0 & -i \\
i & 0
\end{pmatrix}
; \ 
\sigma^3 =
\begin{pmatrix}
1 & 0 \\
0 & -1
\end{pmatrix}.
\end{eqnarray}
We write the amplitudes of a state $\ket{\phi}$ as a column vector $(\product{p}{\phi},\product{n}{\phi})^T$, so that a proton corresponds to $(1,0)^T$, and a neutron $(0,1)^T$.
The inspection of the operator $\bvec\tau \cdot \bvec \pi$, given in Eq.~(\ref{eq:taudotpi}), leads to the identification of
$(\pi_x - i\pi_y)/\sqrt{2}$ with the annihilation of a $\pi_+$ (or with the creation of a $\pi_-$) and $(\pi_x + i\pi_y)/\sqrt{2}$ with the annihilation of a $\pi_-$
(or with the creation of a $\pi_+$).

\section{Lowest order self-energy from the nonrelativistic pion-nucleon Hamiltonian}
\label{app:selfen}
\begin{figure}[!htb]
\centering
\includegraphics[width=0.7\linewidth]{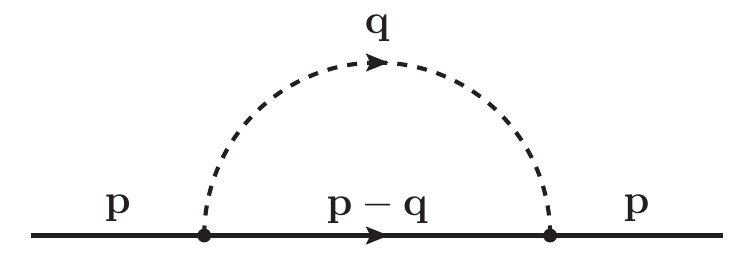}
\caption{Diagram for the lowest order self-energy $\Sigma(E,\bvec{p})$.}
\label{fig:diagram}
\end{figure}

In this calculation we consider only the lowest order interaction term, represented by the diagram of Fig.~\ref{fig:diagram}.

The nonrelativistic propagator for a free nucleon including the
mass counter terms is
\begin{align}
&G(\bvec{x}-\bvec{x}^\prime,t-t^\prime) = \nonumber\\
&=-i \theta(t-t^\prime) \langle 0|N(\bvec{x})e^{-iH(t-t^\prime)}N^\dagger(\bvec{x}')|0\rangle \nonumber\\
&= -i \theta(t-t^\prime) \frac{1}{L^3}\sum_{\bvec{p}} e^{-i\bvec{p}\cdot(\bvec{x}-\bvec{x}^\prime)} e^{-i \left(\frac{p^2}{2M_P}+\beta_K p^2+M_P+\delta M\right)(t-t^\prime)}\nonumber\\
&=  \frac{1}{L^3}\sum_{\bvec{p}} e^{i\bvec{p}\cdot(\bvec{x}-\bvec{x}^\prime)} \int \frac{d\omega}{2\pi} e^{-i\omega(t-t')} G(\bvec{p},\omega)\,.
\end{align}
In the last line we introduced the Fourier transform,
\begin{align}
\label{eq:prop_nucleon}
G(\bvec{p},\omega) &= -i \int_{-\infty}^\infty dt \theta(t) e^{i\left (\omega-\frac{p^2}{2M_K}+\beta_K p^2-M-\delta M\right ) t-\eta t} \nonumber\\
&= \frac{1}{\omega-\frac{p^2}{2M_K}-\beta_K p^2-M_P-\delta M+i\eta},
\end{align}
where $\eta$ is a positive infinitesimal, which was added
to make the integral at the upper limit converges.

The free pion propagator corresponds to that of a free harmonic oscillator
with frequency $\omega_q=\sqrt{q^2+m_\pi^2}$,
\begin{align}
\label{eq:prop_HO}
G_{HO}(\omega) = \frac{1}{\omega^2-\omega_q^2+i\eta}.
\end{align}

Equations (\ref{eq:prop_nucleon}) and (\ref{eq:prop_HO}), together with
standard Feynman diagram rules \cite{fet03}, provide an expression for the self-energy,
\begin{align}
&\Sigma(E,\bvec{p}) = 3i\left (\frac{g_A}{2f_\pi}\right )^2 \frac{1}{L^3}\sum_{\bvec{q}} \int_{-\infty}^\infty \frac{d\omega }{2\pi}
\frac{1}{\omega^2-\omega_q^2+i\eta}  \nonumber\\
&\quad \times \frac{q^2}
{E-\omega
-\left (\frac{1}{2M_P}+\beta_K\right) \left |\bvec{p}-\bvec{q}\right|^2-M_P
-\delta M+i\eta} ,
\end{align}
where the factor of 3 comes from $\bvec \tau \cdot \bvec \tau$ (or the 3 types of hermitian pions). Performing the integral over
$\omega$ yields
\begin{align}
\Sigma(E,\bvec{p}) &= \frac{3}{2}\left (\frac{g_A}{2f_\pi}\right )^2
\frac{1}{L^3}\sum_{\bvec{q}} \frac{1}{\omega_q}\nonumber\\
&\times \frac{q^2}
{E
-\left (\frac{1}{2M_P}+\beta_K\right ) \left |\bvec{p}-\bvec{q}\right|^2-M_P
-\delta M-\omega_q} .
\label{eq:spe}
\end{align}
The single-nucleon spectrum is dictated by the pole of the Green's function,
\begin{equation}
E = \left (\frac{1}{2M_P}+\beta_K\right ) p^2+M_p+\delta M +\Sigma(E,\bvec p)\,.
\end{equation}
We must adjust $\beta_K$ and $\delta M$ so that at small momentum,
$E= M_P+\frac{p^2}{2M_P}$, or
\begin{equation}
\beta_K p^2+\delta M
+\Sigma\left (M_P+\frac{p^2}{2M_P},\bvec p\right)=0\,.
\end{equation}
Expanding in powers of $p$, we find,
\begin{eqnarray}
\label{eq:selflowp}
0 &=& \delta M
-\frac{3}{2}\left (\frac{g_A}{2f_\pi}\right )^2
\frac{1}{L^3}\sum_{\bvec{q}} \frac{q^2}{\omega_q D_q}
\nonumber\\
0 &=& \beta_K
+\beta_K \frac{3}{2}\left (\frac{g_A}{2f_\pi}\right )^2
\frac{1}{L^3}\sum_{\bvec{q}} \frac{q^2}{\omega_q D_q^2}
\nonumber\\
&&
-\left (\frac{1}{M_P}+2\beta_K \right)^2
\frac{1}{2}\left (\frac{g_A}{2f_\pi}\right )^2
\frac{1}{L^3}\sum_{\bvec{q}}\frac{q^4}{\omega_q D_q^3},
\nonumber\\
\end{eqnarray}
where
\begin{equation}
D_q = \delta M 
+\left (\frac{1}{2M_P}+\beta_K\right )q^2 +\omega_q\,.
\end{equation}
Solving these self consistently gives the lowest order values of
$\delta M$ and $\beta_K$. We see from the form above, that the kinetic
mass renormalization is small.

\section{Another choice of the contact interaction}
\label{app:contact}

We also considered
a different functional form for the contact interactions present in
Eq.~(\ref{eq:hqmc}),
the same form as the one used in local chiral EFT potentials~\cite{gez14},
\begin{equation}
\label{eq:deltaR0}
\delta_{R_0}(\bvec r)=\frac{1}{\pi \Gamma(3/4) R_0^3}\exp\left[-(|\bvec r|/R_0)^4\right],
\end{equation}
where $\Gamma$ is the gamma function, and $R_0=1.2$ fm.
We believe that the smeared out function of Eq.~(\ref{eq:deltakc})
is the most appropriate choice for our simulations because
it has the same cutoff as the pion modes, otherwise our
results would depend on multiple cutoffs.
However, to demonstrate that our fitting procedure is compatible
with other QMC simulations,
we also fitted the low-energy constants using
Eq.~(\ref{eq:deltaR0}) for the 
delta function in the contact
terms.
This function should, in principle, 
be modified as in 
Eq.~(\ref{eq:periodic_v}). However,
since it is short-ranged 
compared to the one-pion exchange
potential, for $25 \leqslant L \leqslant 35$ fm we find that
we do not require the potential from the surrounding boxes.
We present our results in Table~\ref{tab:contactR0}.

\begin{table*}[!htb]
\centering
\caption{
Contact parameters using the smeared out delta function of Eq.~(\ref{eq:deltaR0}) for different box sizes, $L$=25, 30, and 35 fm, as a function of the cutoff $\omega_c^s$.
The $\omega_c^s$ are given in MeV, while $C_S$ and $C_T$ are in fm$^2$.}
\begin{tabular}{|c|c|c|c||c|c|c||c|c|c|}
\hline
\multicolumn{1}{|l|}{} &
\multicolumn{3}{c||}{$L$= 25 fm} &
\multicolumn{3}{c||}{$L$= 30 fm} &
\multicolumn{3}{c|}{$L$= 35 fm} \\ \hline
\multicolumn{1}{|l|}{n} &
$\omega_c^s$ & $C_S$ & $C_T$ &
$\omega_c^s$ & $C_S$ & $C_T$ &
$\omega_c^s$ &$C_S$ & $C_T$ \\ \hline\hline
1 & 150.06 & -3.342 & -0.185 & 146.49 & -3.333 & -0.197 & 144.30 & -3.308 & -0.219\\ \hline
2 & 160.61 & -3.409 & -0.140 & 154.06 & -3.372 & -0.165 & 149.98 & -3.308 & -0.219\\ \hline
3 & 166.05 & -3.444 & -0.121 & 158.02 & -3.395 & -0.149 & 152.97 & -3.310 & -0.217\\ \hline
4 & 169.68 & -3.474 & -0.109 & 160.67 & -3.412 & -0.137 & 154.99 & -3.310 & -0.217\\ \hline
5 & 181.85 & -3.579 & -0.085 & 169.67 & -3.466 & -0.108 & 161.88 & -3.312 & -0.215\\ \hline
6 &        &        &        & 177.08 & -3.512 & -0.100 & 167.61 & -3.317 & -0.211\\ \hline
7 &        &        &        & 180.40 & -3.521 & -0.094 & 170.19 & -3.307 & -0.222\\ \hline
8 &        &        &        &        &        &        & 176.06 & -3.316 & -0.221\\ \hline
9 &        &        &        &        &        &        & 180.28 & -3.310 & -0.217\\ \hline
10&        &        &        &        &        &        & 184.20 & -3.361 & -0.151\\ \hline
\end{tabular}
\label{tab:contactR0}
\end{table*}

It is worth mentioning that the chiral potential of
Ref.~\cite{gez14} at LO gives a deuteron binding energy of
$E_d=-2.02$ MeV, which considerably differs from the experimental 
value $-2.23$ MeV. Hence, we found different
values of $C_S$ and $C_T$ than those reported
in \cite{gez14} for the LO potential. 
Additional reasons for this difference are the 
finite volume of the box, and the momentum cutoff that we employ. 
Finally, subleading multiple
pion-exchange contributions, fully accounted for in our 
calculations, appear at NLO in the standard power counting of the 
chiral interaction.

\bibliography{article.bib}
\end{document}